\documentclass{iopart}
\usepackage{epsfig,float,afterpage,amssymb,wrapfig,psfrag}
\newcommand{\be}{\begin{equation}}
\newcommand{\cc}{\mbox{\scriptsize{c}}}
\newcommand{\s}{\mbox{\scriptsize{S}}}
\newcommand{\ii}{\mbox{\scriptsize{i}}}

\newcommand{\ee}{\end{equation}}
\newcommand{\bea}{\begin{eqnarray}}

\newcommand{\host}{\mbox{\ssz{host}}}
\newcommand{\eea}{\end{eqnarray}}

\newcommand{\ssz}{\scriptsize}
\renewcommand{\rmd}{\mbox{d}}
\newcommand{\ei}{\epsilon_{\ii}}
\newcommand{\scrG}{{\cal G}}

\newcommand{\w}{\omega}
\newcommand{\K}{\mbox{\scriptsize{K}}}
\newcommand{\m}{\mbox{\scriptsize{m}}}
\newcommand{\wm}{\w_{\m}}
\newcommand{\wk}{\w_{\K}}
\newcommand{\wl}{\w_{\mbox{\scriptsize{L}}}}
\newcommand{\im}{\mbox{i}}
\newcommand{\subi}{\mbox{\scriptsize{I}}}
\newcommand{\subr}{\mbox{\scriptsize{R}}}

\newcommand{\up}{\uparrow}

\newcommand{\st}{\tilde{\Sigma}}

\newcommand{\down}{\downarrow}
\newcommand{\dn}{\downarrow}
\newcommand{\nb}{\bar{n}}
\newcommand{\mb}{\bar{\mu}}
\newcommand{\eistar}{{\epsilon_{\ii}^*}}
\DeclareMathAlphabet{\bi}{OML}{cmm}{b}{it}
\newcommand{\kk}{\bi{k}}

\newcommand{\nimp}{n_{\mbox{\ssz{imp}}}}

\newcommand{\sts}{\tilde{\Sigma}_{\sigma}}

\renewcommand{\theequation}{\arabic{section}.\arabic{equation}}
\newcommand{\ra}{\rightarrow}

\renewcommand{\Or}{\cal{O}\rm}
\newcounter{saveeqn}
\newcommand{\alpheqn}{\setcounter{saveeqn}{\value{equation}}%
\setcounter{equation}{0}%
\addtocounter{saveeqn}{1}%
\renewcommand{\theequation}{\mbox{\arabic{section}.\arabic{saveeqn}\alph{equation}}}%
}

\newcommand{\reseteqn}{\setcounter{equation}{\value{saveeqn}}%
\renewcommand{\theequation}{\arabic{section}.\arabic{equation}}}

\newcommand{\prb}{{\it Phys. Rev.} B }
\newcommand{\jpcm}{{\it J. Phys.: Condens. Matter}}
\newcommand{\prl}{{\it Phys. Rev. Lett. }}
\newcommand{\pr}{{\it Phys. Rev. }}
\newcommand{\rmp}{{\it Rev. Mod. Phys. }}

\newcommand{\seceq}{\setcounter{equation}{0}}

\begin{document}
\jl{3}
\title[Local quantum phase transition in the pseudogap Anderson model]{Local quantum phase transition in the pseudogap Anderson model: scales, scaling and quantum critical dynamics}
\author{Matthew T Glossop and David E Logan}
\address{Oxford University, Physical and Theoretical Chemistry Laboratory, South Parks Road, Oxford OX1 3QZ, UK}

\begin{abstract}
The pseudogap Anderson impurity model provides a paradigm for
understanding local quantum phase transitions, in this case between
generalised fermi liquid and degenerate local moment phases. Here we develop
a non-perturbative local moment approach to the generic asymmetric
model, encompassing all energy scales and interaction strengths and leading
thereby to a rich description of the problem.
We investigate in particular underlying phase boundaries, the critical
behaviour of
relevant low-energy scales, and single-particle dynamics embodied in the local
spectrum $D(\omega)$. Particular attention is given to the resultant universal
scaling behaviour of dynamics close to the transition in both the GFL and
LM phases, the scale-free physics characteristic of the quantum critical point
itself, and the relation between the two.
\end{abstract}

\maketitle

\section{Introduction}

 The essential physics of a single magnetic impurity coupled to a fermionic
host is embodied at its simplest in the Anderson impurity model (AIM) [1]:
a correlated, non-degenerate impurity with local interaction $U$, hybridizing
to
a non-interacting host band with density of states $\rho_{\mbox{\ssz{host}}}(\omega)$ (for
a comprehensive review see [2]). Yet its simplicity is nominal ---
even for the conventional case of a metallic host, the basic
model for understanding magnetic impurities in metals [1,2] and highly topical
again in the context of quantum dots for example [3,4], or surface atoms
probed by scanning tunneling microscopy (STM) [5].  Here the essential strong
coupling
(large-$U$) behaviour is that of the Kondo effect, characterised by a single
low-energy Kondo scale $\wk$, manifest vividly in the Kondo
resonance in the local single-particle spectrum $D(\omega)$, and rendered no
less
subtle for being by now quite well understood.

There is however one sense in which the metallic AIM is `simple':
it lacks a non-trivial quantum phase transition. The impurity spin is quenched
ubiquitously by coupling to the low-energy degrees of freedom of the metallic
host,
the system is a Fermi liquid for all interaction strengths $U$ and the Kondo
scale,
while exponentially small in strong coupling, never quite vanishes.

That situation changes markedly if the host contains instead a pseudogap
in the vicinity of the Fermi level [6] ($\omega =0$), namely
$\rho_{\host}(\omega) \propto |\omega|^{r}$ with $r>0$ (and $r=0$ corresponding
to the
normal metallic AIM). This pseudogap Anderson impurity model (PAIM)
has been studied extensively in recent years [6-20], particularly for large-$U$
where the low-energy subspace maps onto that of the associated Kondo model,
and from which work much progress and insight has accrued.
The resultant depletion of host states around the Fermi level leads in general
to
the occurrence [6-20] of a quantum phase transition at a finite, $r$-dependent
$U_{\cc}$.
Above the quantum critical point (QCP) a degenerate local moment (LM) phase
arises, with a residual, unquenched impurity spin. For $U < U_{\cc}$ by contrast,
and perturbatively connected to the non-interacting limit ($U=0$), resides a
`strong coupling' or generalised Fermi liquid (GFL) phase in which the
impurity spin is locally quenched and a Kondo effect manifest, reflected in
a low-energy Kondo scale $\wk$.
In the vicinity of the transition each phase is in fact characterised by
a low-energy scale [17-20], here
denoted generically by $\omega_{*}$; corresponding to the Kondo scale
$\omega_{*} \equiv \wk$ for the GFL phase and its counterpart
$\omega_{*} \equiv \wl$ [20] for the LM phase. Most importantly, and
symptomatic
of the underlying transition, the $\omega_{*}$'s vanish as the
QCP is approached.

  Although the PAIM serves in general as a paradigm for understanding local
(boundary) quantum phase transitions [21], many examples of specific systems
with
a power-law density of states have also been discussed. These include various
zero-gap semiconductors [22], one-dimensional interacting systems [23], and the important problem of impurity moments in $d$-wave superconductors [17,18],
notably $Zn$-
and $Li$- doped cuprate materials. Likewise it has recently been shown [19]
that the pseudogap Kondo model exhibits critical local moment fluctuations, and an
associated destruction of the Kondo effect, very similar to those present at
the
local QCP of the Kondo lattice [24] that has been argued to explain the
behaviour of
heavy fermion metals such as CeCu$_{6-x}$Au$_{x}$.

 While great progress has been made in understanding the PAIM [6-20], much
nevertheless
remains to be understood; particularly regarding the quantum critical behaviour
of dynamical properties close to and on either side of the transition, and
precisely \it at \rm the QCP itself. Since the low-energy scale $\omega_{*}$ is arbitrarily small (but non-zero) sufficiently close to the transition, the
frequency
($\omega$) dependence of dynamics should scale in terms of it --- i.e.
exhibit universal scaling as a function of $\omega/\omega_{*}$ --- with
resultant scaling behaviour that will depend on the particular phase
considered, GFL or LM. What precisely is that behaviour for the selected
dynamical property? And how from it does (or can) one understand the scale-free
physics characteristic of the QCP itself, where $\omega_{*} =0$ identically?
These are tough issues, especially from an analytical perspective, and some key
aspects of which have recently been uncovered \it via \rm
numerical renormalization group (NRG) calculations [16,18,19], the local
moment approach [15,16,20] and, for the multichannel pseudogap Kondo model, a
dynamic
large-$N$ approach [17]. Single-particle spectra embodied in the impurity $D(\omega)$
provide an obvious example of dynamics [12, 14-20]; and an important one
given e.g. the direct manifestation in the GFL phase of low-energy Kondo
physics embodied in the Kondo resonance, as well as their more
general relevance to STM and related experimental techniques.

  In this paper we develop a local moment approach (LMA) [25-30,15,20] to the
PAIM
at $T=0$,
for the general case of the particle-hole asymmetric model (noting that the
corresponding Kondo model thus contains both exchange \it and \rm potential
scattering).
Single-particle dynamics form a central focus of the theory albeit,
importantly, that
its intrinsic formulation also enables phase boundaries between the GFL and LM
phases to be determined, as well as the critical behaviour of associated
low-energy scales [15,20]. Although developed originally to describe metallic
Anderson impurity models, at $T=0$ [25-27], finite temperatures [28] and
magnetic
fields [29], the approach is quite general; readily extended
for example to encompass lattice-based models within the framework of dynamical
mean-field theory, such as the periodic Anderson lattice model [30].
It has also been applied previously to and has led to a number of new
predictions
for, the symmetric limit of the PAIM [15,20]; rather successfully, as judged
by subsequent comparison to NRG calculations [16]. The asymmetric PAIM
considered here is however another matter: the symmetric model, recovered
though it is as a particular limit of the present work, has long been
argued [13,17,18] to be a somewhat special case, in some sense divorced from
generic asymmetric behaviour (although our own caveats on the matter will
become
clear in due course).

The LMA, while certainly approximate, is also well
suited to the present problem. Its intrinsically non-perturbative nature
enables
it to handle strong electron correlations and the attendant, central low-energy
scales; while at the same time permitting access to dynamics on all energy
scales,
whether it be on the $\omega_{*}$ scales relevant to universal scaling
behaviour,
or scales on the order of $\omega \sim U$ relevant to the high-energy Hubbard
satellites. Importantly too, and in contrast to difficulties inherent to a
number of
other potential theoretical approaches, the method handles both the GFL and LM
phases
on an essentially equal footing (the LM phase, for example, is certainly not a
trivially decoupled/isolated spin).

The paper is organised as follows. After the requisite background,
we consider in \S 2.1 an obvious question: since the GFL phase is
pertubatively connected to the non-interacting limit, satisfying
as such the Luttinger theorem [31], what is the analogue of the
resultant Friedel sum rule? That is entirely familiar for the
conventional metallic AIM ($r=0$) [2,32], and its counterpart for
the PAIM plays a central role in the present work. In \S 3 another
important issue is considered: what can be deduced on general,
minimalist grounds about the behaviour of the scaling spectra
close to the quantum phase transition; in both the GFL and LM
phases and at the QCP itself? Such considerations are revealing,
and dictate constraints to be met by any credible approximate
theory. They show moreoever that a variety of low-energy scales in
addition to $\omega_{*}$ enter the scaling behaviour of dynamics
--- the quasiparticle weight $Z$ and renormalised level $\epsilon_{\ii}^{*}$
for the GFL phase, together with the latter's spin-dependent counterparts for
the LM phase. Each vanishes as the transition is approached, with distinct
but related critical exponents; but nonetheless enter dynamics in such a way
that
universal scaling remains determined solely by the key low-energy scale
$\omega_{*}$.

 The LMA itself is developed in \S 4, focusing on the notion of
symmetry restoration that is central to the approach [15,20,25-30], together
with
what likewise proves important in dealing with the asymmetric PAIM, namely
self-consistent satisfaction of the Friedel sum rule/Luttinger theorem.
Results arising from the approach are presented in \S s 5-7, beginning
with the critical behaviour of relevant low-energy scales (\S 5.1)
and a determination of representative phase diagrams for the problem (\S 5.2).
Single-particle dynamics are considered in \S 6, on all energy scales
whether universal or non-universal; but with a dominant focus on scaling
spectra in the GFL and LM phases --- on \it all \rm $\omega/\omega_{*}$ scales
--- as well
as the scale-free dynamics arising at the QCP. Finally, we consider in
\S 7 the regime of small $r \ll 1$, obtaining explicit analytic results for the
critical behaviour of low-energy scales, resultant phase boundaries and
single-particle scaling spectra. The paper concludes with a brief
summary/outlook.

\section{Background}
\label{background}
With the Fermi level taken as the energy origin, the Hamiltonian for the
spin-$\frac{1}{2}$ AIM [1,2] is given by
\begin{equation}
\fl \ \ \ \ \ \hat{H}=\sum_{\kk,\sigma}\epsilon_{\kk}\hat{n}_{\kk\sigma}+\sum_{\sigma}(\epsilon_{\ii}+\mbox{$\frac{1}{2}$}U\hat{n}_{\ii -\sigma})\hat{n}_{\ii\sigma}+\sum_{\kk, \sigma}V_{\ii\kk}(c^{\dagger}_{\ii\sigma}c_{\kk\sigma}+c^{\dagger}_{\kk\sigma}c_{\ii\sigma}).
\end{equation}
The first term describes the non-interacting host with density of states
$\rho_{\mbox{\ssz{host}}}(\omega) = \sum_{\kk}\delta(\omega - \epsilon_{\kk})$, while
the second refers to the correlated impurity with bare site-energy $\epsilon_{\ii}$ and on-site interaction $U$. The third term is the one-electron host-impurity
coupling embodied in $V_{\ii\kk}$, taken for convenience as $V_{\ii\kk} = V$
(although that is not required). The particle-hole asymmetry
of the model is specified by the asymmetry parameter [1,27]
\begin{equation}
\eta=1+\frac{2\ei}{U}
\label{eta}
\end{equation}
such that $\eta =0$ corresponds to the particle-hole symmetric case, where
$\epsilon_{i} = -\frac{U}{2}$ and the impurity charge $n = \sum_{\sigma}
\langle \hat{n}_{\ii\sigma} \rangle =1$ for all $U$. In strong coupling, where
the relevant low-energy sector of the AIM maps under a Schrieffer-Wolff
transformation onto a Kondo model containing both exchange ($J$) and
potential ($K$) scattering contributions [2], the asymmetry
$\eta$ is equivalently the ratio of the potential/exchange scattering
strengths: $\eta \equiv 1 -2|\epsilon_{\ii}|/U = K/J$ (see e.g.\ [27]).

  We focus here on the impurity Green function $G(\omega) (\leftrightarrow
G(t) = -\im\langle \hat{T}(c_{\ii\sigma}(t)c^{\dagger}_{\ii\sigma}) \rangle)$, and
hence the local spectrum $D(\omega) = - \frac{1}{\pi}\mbox{sgn}(\omega)\mbox{Im}G(\omega)$.
In addition to intrinsic interest in single-particle dynamics \it per se\rm,
this also provides a direct route to both the location of phase boundaries between
the GFL and LM states [15] and the critical behaviour of associated low-energy
scales, as will be shown in \S s 4,5.

$G(\omega)$ is conventionally expressed as
\begin{equation}
G(\omega) = [\omega^{+} - \epsilon_{\ii} - \Delta(\omega) - \Sigma(\omega)]^{-1}
\end{equation}
with $\omega^{+} = \omega +\im 0^{+}\mbox{sgn}(\omega)$ and $\Sigma(\omega) = \Sigma^{\subr}(\omega)
- \im\mbox{sgn}(\omega)\Sigma^{\subi}(\omega)$ the usual single self-energy; the corresponding
non-interacting Green function is denoted by $g(\omega)$. All effects of
host-impurity coupling at one-electron level are embodied in the hybridization function
$\Delta(\omega) = \sum_{\kk} V_{\ii\kk}^{2} [\omega^{+}-\epsilon_{\kk}]^{-1}$ ($= \Delta_{\subr} -\im\mbox{sgn}(\omega)\Delta_{\subi}(\omega)$). For the PAIM considered,
$\Delta_{\subi}(\omega) = \pi V^{2} \rho_{\mbox{\ssz{host}}}(\omega)$ is given explicitly by
$\Delta_{\subi}(\omega) = \Delta_{0}(|\omega|/\Delta_{0})^{r}\theta(D-|\omega|)$ with
$\Delta_{0}$ the hybridization strength and $D$ the bandwidth ($\theta(x)$ is the
unit step function). That this form is naturally simplified is of course
immaterial to the relevant low-energy physics of the problem in the scaling
regime of primary interest.
We take $\Delta_{0} \equiv 1$ as the basic energy unit in this paper, i.e.\
\begin{equation}
\Delta_{\subi}(\omega) = |\omega|^{r}\theta(D-|\omega|)
\end{equation}
with $r>0$ for the pseudogap AIM (and the limit $r=0$ corresponding to the normal
metallic model). The real part $\Delta_{\subr}(\omega)$ follows from Hilbert transformation
and, for later reference, has the low-$\omega$ behaviour (for all $r \geq 0$)
\begin{equation}
\Delta_{\subr}(\omega) = -\mbox{sgn}(\omega) \left \{ \beta(r)|\omega|^{r} +
\frac{2D^{r}}{\pi (r-1)} \frac{|\omega|}{D} + \Or\left[\left(\frac{|\omega|}{D}\right)^{3}\right]
\right \}
\end{equation}
with $\beta(r) = \mbox{tan}(\frac{\pi}{2}r)$ used throughout. Notice from this that
the wide-band limit $D \rightarrow \infty$, so frequently employed for the
metallic AIM (see \it eg \rm [2]), can be taken for
$r<1$, whereupon $\Delta_{\subr}(\omega) = -\mbox{sgn}(\omega)\beta(r)|\omega|^{r}$ for
\it all \rm $\omega$. We consider  $0 < r<1$ in this paper,
which range is assumed unless stated to the contrary; and
for that reason will employ the wide-band limit. This is simply a matter of
convenience, is again irrelevant to the key low-energy behaviour of the system
and can be relaxed trivially.

  The impurity charge $n$ is of course related to the local spectrum
by $n =2\int^{0}_{-\infty} \rmd \omega \ D(\omega)$. The `excess' charge $\nimp$  will also prove important to the present work, for reasons given in \S2.1.
This is defined as the change
in the number of electrons of the entire system due to addition of the impurity
to the host, and is given generally by
\begin{equation}
\nimp = 2 \int^{0}_{-\infty} \rmd\omega \ \Delta\rho(\omega).
\end{equation}
Here $\Delta\rho(\omega)$ ($=\rho(\omega) - \rho_{\mbox{\ssz{host}}}(\omega)$) is the excess
density of states, itself given by (see e.g.\  [2])
\begin{equation}
\Delta\rho(\omega) = -\frac{1}{\pi} \mbox{sgn}(\omega) \mbox{Im} \left \{
G(\omega)\left[1- \frac{\partial\Delta(\omega)}{\partial\omega}\right] \right \}
\end{equation}
such that $\nimp$ is related to $n$ by
\begin{equation}
\nimp = n - \frac{2}{\pi} \mbox{Im} \int^{0}_{-\infty} \rmd\omega
\ G(\omega) \frac{\partial\Delta(\omega)}{\partial\omega}.
\end{equation}

  From the above, for specified gap index $r$, the PAIM is characterised by any two of
the `bare' parameters $U$, $\epsilon_{\ii}$ and $\eta = 1 + 2\epsilon_{\ii}/U$
(together if appropriate with the bandwidth $D$). In practice, as now explained,
we choose to consider $\eta$ and one or other of $U$ and $\epsilon_{\ii}$ as
the basic parameters.
For the $r=0$ metallic AIM we have recently considered [26,27] the
scaling behaviour of $D(\omega)$ arising in the strong coupling Kondo
regime, where $\epsilon_{\ii} < 0$ such that $|\epsilon_{\ii}| \gg 1$ and
$U-|\epsilon_{\ii}| \gg 1$. Here the Kondo scale $\wk$ is exponentially
small. For fixed asymmetry $\eta$ [27], $D(\omega)$ then
exhibits scaling in terms of $\tilde{\omega} = \omega/\omega_{\K}$. That is,
on fixing $\eta = 1-2|\epsilon_{\ii}|/U$ and progressively increasing the
interaction $U$, $D(\omega)$ is universally dependent on
$\tilde{\omega} = \omega/\omega_{\K}$ alone, and otherwise independent of
the bare model parameters (which themselves determine the Kondo scale). The
important point here is that this arises  for {\it fixed} asymmetry
$\eta \equiv K/J$ (and in this case only for fixed $\eta$, i.e.\ universal scaling does not arise upon increasing $U$ for fixed $|\ei|\gg 1$ [27]).  In consequence, a family of scaling spectra arises, one for each
asymmetry $\eta$. This is not widely appreciated (although in the
related context of d.c.\ transport it was understood early by Wilson [33]),
but that it must arise is nearly obvious. For if the scaling spectra for
the generic asymmetric AIM were $\eta$-independent, they would coincide with
that for the p-h symmetric model $\eta =0$ ($=K/J$) and as such be strictly
symmetric in $\omega$ about the Fermi level $\omega =0$; the implausibility
of which is at least intuitive.

  For the pseudogap model considered here, the  quantum
phase transition between GFL and LM phases is naturally characterised by
a low-energy scale $\omega_{*}$ --- the Kondo scale $\wk$ for the GFL
phase and its counterpart $\wl$ [20] (discussed in \S 6ff) for the LM phase
 --- that vanishes
as $U \rightarrow U_{\cc}(r)\pm$ and the transition is approached. In consequence
the spectra likewise exhibit scaling in terms of $\tilde{\omega} = \omega/\omega_{*}$
in the vicinity of the transition. As discussed in \S 6ff, we again find  this
scaling to be $\eta$-dependent.  For this reason, and to enable continuity as $r\ra 0$ to the above scaling behaviour for the metallic AIM, we choose to regard $\eta$ and one or other of $U$ and $\epsilon_{\ii}$ as the
basic model parameters. The minimal range of $\eta$ that need be considered is
also readily established, for under a general p-h transformation of the Hamiltonian
it is straightforward to show that $\epsilon_{\ii} \rightarrow -[\epsilon_{\ii} +U]$
and $D(\omega; \epsilon_{\ii})= D(-\omega; -[\epsilon_{\ii}+U])$, hence
$n(\epsilon_{\ii}) = 2-n(-[\epsilon_{\ii}+U])$ for the impurity charge (and likewise
for the excess charge $\nimp$). From this it follows that only
$\epsilon_{\ii} \geq -\frac{1}{2}U$ need
be considered, corresponding to $\eta \geq 0$ and which range is assumed from now
on.

  To capture the underlying phase transition intrinsic to the PAIM, an inherently
non-perturbative approach is obviously required. That poses severe and, to our
knowledge, insurmountable problems for any approach based directly on the conventional
single self-energy $\Sigma(\omega)$. But such an approach is not obligatory,
$\Sigma(\omega)$ merely being defined by the Dyson equation implicit in equation (2.3)
for $G(\omega)$. For the doubly degenerate LM phase in fact, a two-self-energy (TSE)
description would be physically far more natural (as is obvious by considering
 e.g.\ the atomic limit, an extreme case of the LM phase). Here the rotationally
invariant $G(\omega)$ is expressed as
\begin{equation}
G(\omega) = \mbox{$\frac{1}{2}$}[G_{\uparrow}(\omega) + G_{\downarrow}(\omega)]
\end{equation}
with the $G_{\sigma}(\omega)$ given by
\begin{equation}
G_{\sigma}(\omega) = [\omega^{+} - \epsilon_{\ii} - \Delta(\omega)
-\tilde{\Sigma}_{\sigma}(\omega)]^{-1}
\end{equation}
in terms of self-energies $\tilde{\Sigma}_{\sigma}(\omega)$
($= \tilde{\Sigma}^{\subr}_{\sigma}(\omega) -
\im\mbox{sgn}(\omega)\tilde{\Sigma}^{\subi}_{\sigma}(\omega)$) that are in general distinct and
spin-dependent. The TSE description is in general a necessity and not a luxury for
the LM phase. It is precisely this framework that underlies the LMA [25-27] where,
as explained further in \S 4, it is employed for both the LM and GFL phases and
in consequence is able to capture simultaneously \it both \rm phases and hence
the transition between them. The conventional single self-energy can of course
be obtained as a byproduct of the TSE description.
It follows from direct comparison of equations (2.3,2.9,2.10) and is given for later use by
\begin{equation}
\Sigma(\w)=\mbox{$\frac{1}{2}$}\left[\tilde{\Sigma}_{\up}(\w)+\tilde{\Sigma}_{\down}(\w)\right]+\frac{\left[\frac{1}{2}(\tilde{\Sigma}_{\up}(\w)-\tilde{\Sigma}_{\down}(\w))\right]^2}{g^{-1}(\w)-\frac{1}{2}(\tilde{\Sigma}_{\up}(\w)+\tilde{\Sigma}_{\down}(\w))}
\label{2to1}
\end{equation}
with $g(\omega) = [\omega^{+} -\epsilon_{\ii} - \Delta(\omega)]^{-1}$ the
non-interacting propagator.

  Before turning to the LMA itself we consider two important questions. What can
be deduced on general grounds about the behaviour of the GFL phase? That
is considered in \S 2.1. Second, as considered in \S3, what may be inferred
 via  general scaling arguments about the behaviour of the scaling
spectra close to the phase transition; in both the GFL and LM phases and
at the quantum critical point (QCP) itself, where the low-energy scale
$\omega_{*} =0$ identically?

\subsection{GFL phase: Friedel sum rule}
The GFL phase is perturbatively connected to the non-interacting limit,
which in essence defines that phase. This adiabatic continuity requires that the Luttinger integral vanish [2,31], i.e.\ that
\begin{equation}
I_{\mbox{\ssz{L}}} = \mbox{Im}\int^{0}_{-\infty} \rmd\omega \ \frac{\partial\Sigma(\omega)}{\partial\omega}\ G(\omega) = 0.
\end{equation}
From this, using only $\Sigma^{\subi}(0)=0$, the Friedel sum rule follows
directly. Given explicitly by
\begin{equation}
\nimp = 1 - \frac{2}{\pi} \mbox{tan}^{-1}\left(\frac{\epsilon_{\ii}^{*}}{\Delta_{\subi}(0)}\right)
\end{equation}
it relates the excess charge $\nimp$ to the `renormalised level' $\epsilon_{\ii}^{*}$,
itself defined by
\begin{equation}
\epsilon_{\ii}^{*} = \epsilon_{\ii} + \Sigma^{\subr}(0).
\end{equation}
For the $r=0$ metallic AIM, where
$\Delta_{\subi}(0) = \Delta_{0}$ ($\equiv 1$), this gives the Friedel sum rule in
the familiar form $\epsilon_{\ii}^{*} = \mbox{tan}[\frac{\pi}{2}(1-\nimp)]$ [2]. For the
PAIM by contrast, $\Delta_{\subi}(0) =0$; and in consequence $\nimp$ is restricted
to the integer values
\begin{equation}
 \nimp=
\left\{\begin{array}{ccc}
0 &\ \ \ : \ \ \ & \eistar>0 \\
2&\ \ \ : \ \ \ & \eistar<0\\
\end{array}
\right.\\
\label{nimpinteger}
\end{equation}
according to whether the renormalised level lies above or below the Fermi level.
This behaviour, which we reiterate is a consequence of adiabatic continuity and
the Luttinger theorem, is physically natural: one expects intuitively that the
change in number of electrons due to addition of the impurity should be integral.
It is in fact the more familiar metallic AIM, where from equation (2.13) above $\nimp$
is in general non-integral, that is the exception; reflecting in that case the
existence of a finite density of conduction band states arbitrarily close to
the Fermi level.

  The above conclusions may be checked order-by-order in perturbation
theory (PT) about the non-interacting limit (as we have confirmed explicitly
at simple first and second order levels): the resultant $\nimp$ calculated
from spectral integration (equation (2.6)) indeed satisfies equation (2.15), and $I_{\mbox{\ssz{L}}} =0$.
However the Friedel sum rule and hence Luttinger theorem is \it not \rm satsified
by approximate theories in general, save for the p-h symmetric case where it
is guaranteed by symmetry (and for which $\nimp =1$ regardless of whether
the GFL or LM phase is considered). For the metallic AIM that is well known
to be the case for the non-crossing approximation (NCA) [34], and is true even for second order PT
employing Hartree renormalised propagators [35], usually but loosely referred to as second order PT: the resultant $\nimp$ determined by spectral integration
does not in general concur with that implied by the Friedel sum rule (whether
for the normal metallic model or the PAIM). It is however obviously desirable
that the Friedel sum rule be satisfied if possible, and in \S 4.1 we incorporate
it straightforwardly within the LMA.

\seceq
\section{Scales and scaling: general considerations}
  Close to and on either side of the quantum phase transition intrinsic to
the PAIM, the problem is characterised by a single low-energy scale denoted
generically by $\omega_{*}$ that vanishes as the transition is approached,
$U \rightarrow U_{\cc}(r)\pm$ [17-20]. Given this
we now use general scaling arguments to make a number of deductions about the
scaling spectra, in both the GFL and LM phases and at the quantum critical point (QCP)
itself (where $\omega_{*} =0$). We take it for granted in the following
that the QCP is continuously accessible from either the GFL or LM phases as
$\omega_{*} \rightarrow 0$. Although natural, implicitly assumed in NRG studies [18,19]
and found also to arise within the LMA [20], this remains strictly an assumption
(without which essentially no deductions can be made about the QCP).

  The spectrum is denoted in this section by $D(U;\omega)$ with the $U$-dependence
temporarily explicit (and the $\eta = 1 + 2\epsilon_{\ii}/U$ dependence implicit).
As the transition is approached, $u = |1-U/U_{\cc}(r)| \rightarrow 0$ and the low-energy
scale vanishes, as
\begin{equation}
\omega_{*} = u^{a}
\end{equation}
with exponent $a$. $D(U;\omega)$ can now be expressed generally in the scaling form $\pi
D(U;\omega) = u^{-ab}\Psi_{\alpha}(\omega/u^{a})$ in terms of two exponents
$a$ and $b$ (and with $\alpha =$ GFL or LM denoting the phase), i.e.\ as
\begin{equation}
\pi \omega_{*}^{b} D(U;\omega) = \Psi_{\alpha}(\omega/\omega_{*})
\end{equation}
with the exponent $a$ eliminated and the $\omega$-dependence encoded solely in
$\tilde{\omega} = \omega/\omega_{*}$.  Equation (3.2) simply embodies the universal
scaling behaviour of the single-particle spectrum close to the transition. Since the QCP
must be `scale-free' (independent of $\omega_{*}$), it follows from equation (3.2) that the
leading $x \rightarrow \infty$ behaviour of
$\Psi_{\alpha}(x)$ is independent of $\alpha$ and given by
\begin{equation}
\Psi_{\alpha}(\omega/\omega_{*}) \stackrel{\omega/\omega_{*} \rightarrow
\infty}{\sim}
C(r,\eta)(|\omega|/\omega_{*})^{-b}
\end{equation}
with $C(r,\eta)$ a constant;
such that precisely at the QCP ($\omega_{*} =0$ identically)
\begin{equation}
\pi D(U_{\cc};\omega) = C(r,\eta) |\omega|^{-b}.
\end{equation}
The exponent $b$ thus governs the $\omega$-dependence of the QCP spectrum, and is
independent of the phase $\alpha$. In physical terms equation (3.3) embodies the fact that
the \it large \rm $\tilde{\omega}$ asymptotic `tail' behaviour of the scaling spectra
in \it either \rm phase coincides with that of the QCP. More generally of course, the
$\tilde{\omega}$-dependence of the scaling spectra $\Psi_{\alpha}(\tilde{\omega})$
will differ for each phase.

  For the p-h symmetric PAIM ($\eta =0$) it is readily seen that the exponent
$b=r$, using equation (3.2) and the fact [14] that the leading low-$\omega$ behaviour of
$D(U;\omega)$ in the GFL phase is entirely unrenormalised from the non-interacting
limit and given by
\begin{equation}
\pi D(U;\omega) \stackrel{|\omega| \rightarrow 0}{\sim}
\mbox{cos}^{2}(\frac{\pi}{2}r) |\omega|^{-r} \ \ \ : \ \ \ \eta =0
\end{equation}
We add that the resultant QCP behaviour $\pi D(U_{\cc};\omega) = C(r;0)|\omega|^{-r}$
is precisely as found in an LMA study of the symmetric PAIM [20], and that the
exponent $b=r$ is likewise consistent with NRG calculations for the GFL phase [16].

  The behaviour equation (3.5) is exact [14-16], arising because $\Sigma(\omega)$ vanishes
more rapidly than the hybridization ($\propto |\omega|^{r}$), which itself reflects
adiabatic continuity to the non-interacting limit. That is manifest also in the
limiting low-frequency `quasiparticle form' for the impurity Green function in the GFL
phase. This follows from a low-$\omega$ expansion of the single self-energy
\begin{equation}
\Sigma(\omega) \sim \Sigma^{\subr}(0) - \left(\frac{1}{Z} -1\right)\omega
\end{equation}
with $Z=[1-(\partial\Sigma^{\subr}(\omega)/\partial\omega)_{\omega =0}]^{-1}$ the
quasiparticle weight which likewise vanishes as $U \rightarrow U_{\cc}(r)-$; and
with quasiparticle damping embodied in $\Sigma^{\subi}(\omega)$ neglected as usual [2],
as is readily justified perturbatively.
 For the symmetric PAIM, where $\epsilon_{\ii}
= -U/2$ and the impurity charge $n=1$, $\Sigma^{\subr}(0)$ is given trivially by the
Hartree contribution $\frac{1}{2}U n = \frac{1}{2}U$, whence the renormalised
level $\epsilon_{\ii}^{*} = \epsilon_{\ii} + \Sigma^{\subr}(0)$ vanishes by symmetry. From
equations (2.3) and (3.6) the resultant quasiparticle behaviour is then given in scaling
form by
\begin{equation}
\pi \omega_{*}^{r} D(U;\omega) \sim \frac{|\tilde{\omega}|^{r}}
{(k_{1}\tilde{\omega} + sgn(\omega)\beta(r)|\tilde{\omega}|^{r})^{2} +
|\tilde{\omega}|^{2r}} \ \ \ : \ \ \ \eta =0
\end{equation}
from which equation (3.5) is recovered as $|\tilde{\omega}| \rightarrow 0$; and where the
Kondo scale $\omega_{*}$ and quasiparticle weight $Z$ are related by
\begin{equation}
\omega_{*} = [k_{1}Z]^{\frac{1}{1-r}}
\end{equation}
(which relation we add is indeed found within the LMA [15]).

  For the general asymmetric PAIM ($\eta \neq 0$) we know of no simple argument
to infer the exponent $b$, but NRG calculations are again consistent with
$b=r$ [18] (as one might expect from continuity to the symmetric limit). For the
asymmetric model the renormalised level $\epsilon_{\ii}^{*}$ will in general be
non-vanishing, and the quasiparticle form appropriate to the GFL phase is
then given by
\begin{equation}
\pi\omega_{*}^{r} D(U;\omega) \sim \frac{|\tilde{\omega}|^{r}}
{(k_{1}\tilde{\omega} - (\epsilon_{\ii}^{*}/{\omega_{*}^{r}}) +
\mbox{sgn}(\omega)\beta(r)|\tilde{\omega}|^{r})^{2} + |\tilde{\omega}|^{2r}}.
\end{equation}
Granted $b=r$ it follows from this that the renormalised level $\epsilon_{\ii}^{*}$
must also vanish as $U \rightarrow U_{\cc}(r)-$ and must do so at least as rapidly
as $\omega_{*}^{r}$,  i.e.\
\begin{equation}
|\epsilon_{\ii}^{*}| \sim u^{a_{1}}
\end{equation}
with exponent $a_{1} \geq ar$. The GFL scaling spectrum $\Psi_{\mbox{\ssz{GFL}}}(\tilde{\omega})
= \pi\omega_{*}^{r}D(U;\omega)$ is of course obtained by considering in general
finite $\tilde{\omega} = \omega/\omega_{*}$, in the formal limit $\omega_{*}
\rightarrow 0$. From equation (3.9) there are then two \it a priori \rm possibilities for
its low-$\tilde{\omega}$ behaviour. (a) Either $a_{1} > ar$ and thus
$\epsilon_{\ii}^{*}/\omega_{*}^{r} \rightarrow 0$; whence from equation (3.9) the leading
low-$\tilde{\omega}$ behaviour of the asymmetric GFL scaling spectrum would coincide
with that of the symmetric limit equation (3.7), namely $\omega_{*}^{r}D(U;\omega)
\propto |\tilde{\omega}|^{-r}$. Or (b) $a_{1} = ar$ and the renormalised level
vanishes precisely as $\omega_{*}^{r}$,
\begin{equation}
\omega_{*} = [|\epsilon_{\ii}^{*}|/k]^{\frac{1}{r}}
\end{equation}
with $k$ a constant. In this case the leading low-$\tilde{\omega}$ behaviour of the
scaling spectrum is $\Psi_{\mbox{\ssz{GFL}}}(\tilde{\omega}) \propto |\tilde{\omega}|^{r}$, and the
quasiparticle form equation (3.9) implies an asymmetric double-peak spectrum straddling the
Fermi level. It is precisely this characteristic behaviour that is observed in NRG
calculations [18] (see also \S 6.1 and figure 11 below), from which we conclude that $\omega_{*}$ and
the renormalised level $\epsilon_{\ii}^{*}$ should indeed be related by equation (3.11). NRG
calculations for $\frac{1}{2} < r \leq 1$ [18] are moreover consistent with an exponent
of $a = \frac{1}{r}$ for the vanishing of the low-energy scale $\omega_{*}$ ($\propto T^{*}$ in the notation of [18]); implying in turn that the exponent $a_{1} =1$. The behaviour just
inferred will be considered within the LMA in \S 5,6. We would however
add that it gives by no means the whole story, for example the quasiparticle form is
by definition confined to asymptotically low frequencies $\tilde{\omega}$ and
essentially nothing can be deduced from it regarding the general form of
$\Psi_{\mbox{\ssz{GFL}}}(\tilde{\omega})$.

  The latter remarks have focused on the GFL phase. But the renormalised level
$\epsilon_{\ii}^{*}$ will in general be non-zero in the LM phase as well, with the
leading low-$\omega$ behaviour of the spectrum thus given from equation (2.3) by $\pi D(U;\omega)
\propto |\omega|^{r}/|\epsilon_{\ii}^{*}|^{2}$ (assuming only that $\Sigma(\omega)$
vanishes no less rapidly than the hybridization as $|\omega| \rightarrow 0$). For this
behaviour to be part of the scaling spectrum in the LM regime, and because the
exponent $b$ ($=r$) of equation (3.2) is independent of the phase, it follows that the
low-energy scale $\omega_{*}$ is again related to the renormalised level by
equation (3.11); and hence that $|\epsilon_{\ii}^{*}|$ must vanish as the transition is
approached from \it either \rm phase. This too will be considered within the LMA in \S 5.1.

\seceq
\section{Local moment approach}

  In this section we develop a local moment approach to the PAIM for arbitrary
asymmetry $\eta$. For the particular case of p-h symmetry ($\eta =0$), it reduces to the approach given hitherto in [15]. Further details can be found e.g. in [15,27].

\subsection{Basics and symmetry restoration}

  There are three essential elements to the LMA. First that local moments (`$\mu$'),
regarded as the primary effect of electron interactions [1], are introduced explicitly
from the outset. The starting point is thus simple static mean-field (MF) i.e.\
unrestricted Hartree-Fock (HF), containing in general two, \it degenerate \rm broken
symmetry MF states. Notwithstanding the severe limitations of MF by itself (see e.g.\
[15,25,27]), it can
nonetheless be used as the starting point for a successful many-body theory [25-30,15,20].
To that end the LMA employs the two-self-energy
description mentioned in \S 2, that follows naturally from the underlying two MF
saddle points. The local propagator is thus expressed in the form equations (2.9,10), with
the two self-energies $\tilde{\Sigma}_{\sigma}(\omega) \equiv
\tilde{\Sigma}_{\sigma}[\{\cal{G}_{\sigma}\}]$
$(\sigma = \uparrow,\downarrow)$ built diagrammatically from the MF propagators
(themselves denoted by $\cal{G}_{\sigma}(\omega)$). The self-energies are conveniently
separated as
\begin{equation}
\tilde{\Sigma}_{\sigma}(\omega) = \tilde{\Sigma}^{0}_{\sigma} +
\Sigma_{\sigma}(\omega)
\end{equation}
into a purely static contribution ($\tilde{\Sigma}^{0}_{\sigma}$, figure 1 below) that
alone would be retained
at pure MF level, together with the crucial dynamical piece $\Sigma_{\sigma}(\omega)$
that contains in particular the correlated electron spin-flip physics intrinsic
to both the GFL and LM phases.

 For the GFL phase the third key notion behind the LMA, as now discussed for the PAIM, is that
of symmetry restoration [26-30]: self-consistent restoration of the symmetry broken at
pure MF level, and hence recovery in particular of the low-frequency quasiparticle form
(\S 3) for $G(\omega)$ that is symptomatic of
adiabatic continuity to the non-interacting limit. This is given explicitly by
\begin{equation}
G(\omega) \sim \left[\frac{\omega}{Z} - \epsilon_{\ii}^{*} + \mbox{sgn}(\omega)\beta(r)|\omega|^{r}
+\im\mbox{sgn}(\omega)|\omega|^{r}\right]^{-1}
\end{equation}
(from which equation (3.9) above follows directly). Equation (4.2) reflects the fact that as
$|\omega| \rightarrow 0$ the single self-energy $\Sigma^{\subi}(\omega)$ decays to zero
\it more \rm rapidly than the hybridization ($\propto |\omega|^{r}$), so that
neglect of quasiparticle damping is justified and the leading
low-$\omega$ behaviour of $G(\omega)$ amounts in general to a simple renormalisation
of the non-interacting limit (\S 3). The question then is under what conditions on the
\it two-self-energies \rm $\{\tilde{\Sigma}_{\sigma}(\omega) \}$ does such behaviour
arise? In parallel with the analysis of [27] this is readily answered via a simple
low-$\omega$ expansion of the $\tilde{\Sigma}_{\sigma}(\omega)$'s, namely
\begin{equation}
\tilde{\Sigma}_{\sigma}^{\subr}(\omega) \sim \tilde{\Sigma}_{\sigma}^{\subr}(0) -\left[\frac{1}{Z_{\sigma}} -1\right]\omega
\end{equation}
for the real parts, with $Z_{\sigma} = [1-(\partial\Sigma^{\subr}_{\sigma}(\omega)/
\partial\omega)_{\omega=0}]^{-1}$ thus defined;
together with
\begin{equation}
\tilde{\Sigma}^{\subi}_{\sigma}(\omega) \equiv \Sigma^{\subi}_{\sigma}(\omega) \sim
a_{\sigma}|\omega|^{m}
\end{equation}
for the imaginary part, with $m>r$. This form is in fact guaranteed from
the diagrams (\S 4.3)
for $\Sigma_{\sigma}(\omega) \equiv \Sigma[\{{\cal{G}}_{\sigma} \}]$
with $m = 2+3r$ (which concurs with simple perturbation theory in $U$).
Equations (4.3) \& (4.4) are now used in equations (2.11) \& (2.3) to determine the resultant low-$\omega$
behaviour of $\Sigma(\omega)$ and hence $G(\omega)$. From this one finds simply
that the requisite condition for the quasiparticle form equation (4.2) to arise is
\begin{equation}
\tilde{\Sigma}^{\subr}_{\uparrow}(\omega =0) = \tilde{\Sigma}^{\subr}_{\downarrow}(\omega =0).
\end{equation}
If equation (4.5) is \it not \rm satisfied, then the resultant leading low-$\omega$
behaviour of $\Sigma^{\subi}(\omega)$ is $ \propto |\omega|^{r}$ which, vanishing
as rapidly as the hybridization, vitiates the quasiparticle form equation (4.2).

 Equation (4.5) is of course the familiar symmetry restoration condition for the GFL phase
likewise found
for the $r=0$ metallic AIM [26,27], and for the symmetric PAIM [15] (where it
reduces to $\tilde{\Sigma}^{\subr}_{\sigma}(0) =0$ independently of spin $\sigma$,
due to p-h symmetry). As in previous work [25-30,15,20], it amounts in practice (see also \S 4.2)
to a self-consistency condition for the local moment $|\mu|$ entering the MF
propagators $\{ {\cal{G}}_{\sigma}(\omega) \}$ from which the two self-energies
$\{ \tilde{\Sigma}_{\sigma}(\omega) \}$ are diagrammatically constructed. This
generates directly [25,27] a low-energy scale, seen as a sharp
resonance (at $\omega = \wm$) in the spectrum of transverse spin
excitations \mbox{Im}$~\Pi^{+-}(\omega)$ (\S 4.2); which is non-zero throughout the
GFL phase and sets the timescale $\tau \sim h/\omega_{m}$ for
symmetry restoration and hence local `Kondo singlet' formation. This is the Kondo scale, in terms of which single-particle dynamics are found to scale in \S 6; and with $\wm\equiv\w_{*}=\wk$ in the notation employed above (low-energy scales are of course in general defined only up to an arbitrary constant).

Imposition of
symmetry restoration as a self-consistent constraint thus again underlies the
LMA for the GFL phase. If equation (4.5) is satisfied then (i) the requisite low-energy
quasiparticle form is recovered, as above; (ii) all self-energies coincide at the
Fermi level, namely $\Sigma^{\subr}(0) = \tilde{\Sigma}^{\subr}_{\sigma}(0)$ for either
$\sigma$ (whence the renormalised level equation (2.14) is given equivalently by $\epsilon_{\ii}^{*} = \epsilon_{\ii} + \tilde{\Sigma}^{\subr}_{\sigma}(0)$); and
(iii) the quasiparticle weight $Z$ is related to the $\{Z_{\sigma} \}$ by
$Z^{-1} = \frac{1}{2}(Z^{-1}_{\uparrow} + Z^{-1}_{\downarrow})$. It is moreover
the limits of solution to the symmetry restoration condition that delineate the
phase boundaries between the GFL and LM phases (\S 5.2), as for the symmetric PAIM
considered in [15]
(for $U > U_{\cc}(r)$ in the doubly degenerate LM phase itself, discussed explicitly
in \S 4.3, symmetry is not of course restored).

  While solution of the symmetry restoration condition equation (4.5) ensures
correct recovery of the quasiparticle form for the GFL phase, it does not
in general lead to the Friedel sum rule (equation (2.15)) being satisfied. We have
recently developed a strategy for metallic AIMs [27] under which both symmetry
restoration and the Friedel sum rule may be simultaneously satisfied within a
two-self-energy framework. That approach may be modified straightforwardly
to encompass the PAIM, as now sketched (and regardless of specific details
of the $\tilde{\Sigma}_{\sigma}(\omega) \equiv \tilde{\Sigma}_{\sigma}[\{
{\cal{G}}_{\sigma} \}]$ which are considered separately in \S 4.2).

\subsubsection{Algorithm}

  The two-self-energies are functionals of the MF propagators
$\{ {\cal{G}}_{\sigma}(\omega) \}$ [27], given by
\begin{equation}
{\cal{G}}_{\sigma}(\omega) = [\omega^{+} - e_{\ii} +\sigma x - \Delta(\omega)]^{-1}
\end{equation}
where $x=\frac{1}{2}U|\mu|$ and $e_{\ii}$ differ in general from their
MF values $x_{0} = \frac{1}{2}U|\mu_{0}|$ and $e_{\ii 0} = \epsilon_{\ii}
+\frac{1}{2}Un_{0}$ (with $|\mu_{0}|$ and $n_{0}$ the local moment and charge
obtained self-consistently at pure MF level). The $\{ \tilde{\Sigma}_{\sigma}(\omega)
\}$ thus naturally depend on both $e_{\ii}$ and $x$:
$\tilde{\Sigma}_{\sigma}(\omega) = \tilde{\Sigma}_{\sigma}(\omega;e_{\ii},x)$;
and, via equations (2.9) \& (2.10), so too does $G(\omega) = G(\omega;e_{\ii},x)$ and hence
$\nimp = \nimp(e_{\ii},x)$. With such dependence made temporarily explicit,
the symmetry restoration condition and Friedel sum rule (\S 2.1) for the GFL phase read
\begin{equation}
\tilde{\Sigma}_{\uparrow}^{\subr}(0;e_{\ii},x) = \tilde{\Sigma}^{\subr}_{\downarrow}(0;e_{\ii},x)
\end{equation}
and
\begin{equation}
\fl \ \ \ \nimp(e_{\ii},x) = \frac{2}{\pi}\mbox{Im} \int^{0}_{-\infty}\rmd\omega\
G(\omega)\left[1- \frac{\partial\Delta(\omega)}{\partial\omega}\right] =
\left \{\begin{array}{ccc} 0 & : & \epsilon_{\ii}^{*}(e_{\ii},x)>0 \\
2 & : & \epsilon_{\ii}^{*}(e_{\ii},x) <0
\end{array} \right. .
\end{equation}
These equations are together sufficient to determine both $x$ and $e_{\ii}$.
In practice, for any given $U$, it is natural to solve equations (4.7) \& (4.8) by fixing
$e_{\ii}$ and determining $\epsilon_{\ii}$, in the following way: (i) For given
$e_{\ii}$ solve the symmetry restoration condition equation (4.7) for $x=\frac{1}{2}U|\mu|$
(and hence $|\mu|$).
(ii) The self-energies $\{ \tilde{\Sigma}_{\sigma}(\omega;e_{\ii},x) \}$ are now
completely determined, and the bare $\epsilon_{\ii}$ entering $G(\omega)$
(equations (2.9) \&(2.10)) is then varied until the Friedel sum rule equation (4.8) is satisfied.
All quantities are now known without looping and the procedure is a rapid numerical
exercise. Alternatively and equivalently, if one wishes to specify a bare
$\eta = 1 + 2\epsilon_{\ii}/U$ (or $\ei$) from the outset, the procedure may be repeated
by varying $e_{\ii}$ until equation (4.8) is satisfied.

 Further, following [27] it is straightforward to show that $e_{\ii} \rightarrow
-e_{\ii}$ corresponds to the particle-hole transformation mentioned in \S 2, under
which $\epsilon_{\ii} \rightarrow -[\epsilon_{\ii} +U]$ (i.e.\ $\eta\ra -\eta$), $\nimp \rightarrow
2-\nimp$ and $D(\omega) \rightarrow D(-\omega)$. As for the metallic AIM [27]
we can thus focus solely on $e_{\ii} \geq 0$. The case of $e_{\ii} =0$ is of course
the particle-hole symmetric model [15,25] where the Luttinger integral, equation (2.12),
vanishes by symmetry, and $\nimp = 1$ for all $U$; in this case step (ii) above
then drops out of the problem and the strategy is as presented hitherto in [15].

\subsection{Self-energies}

  We now specify the particular approximation employed for the dynamical
contribution $\Sigma_{\sigma}(\omega)$ to the two-self-energies. Equation (4.1)
for $\tilde{\Sigma}_{\sigma}(\omega)$ is given explicitly by
\begin{equation}
\tilde{\Sigma}_{\sigma}(\omega) = \mbox{$\frac{1}{2}$}U(\bar{n} -\sigma |\bar{\mu}|)
+ \Sigma_{\sigma}(\omega)
\end{equation}
\begin{figure}
\begin{center}
\psfrag{sigma}[bc][bc]{{\large \bf $\sts^0 \equiv $}}
\epsfig{file=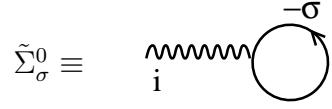,width=4cm}
\caption{Static bubble diagram contribution to $\tilde{\Sigma}_{\sigma}(\omega)$. The wavy line denotes $U$, and the solid line the broken symmetry MF propagator.}
\label{static}
\end{center}
\end{figure}
where the first term is the static bubble diagram (figure 1),
$\tilde{\Sigma}^{0}_{\sigma} = U\int^{0}_{-\infty}\rmd\omega \
D^{0}_{-\sigma}(\omega;e_{\ii}, x)$ with $|\bar{\mu}| \equiv |\bar{\mu}(e_{\ii},x)|$
and $\bar{n} \equiv \bar{n}(e_{\ii},x)$ given by
\alpheqn
\be
|\mb|=\sum_{\sigma}\sigma\int_{-\infty}^0\rmd\w\ D_{\sigma}^0(\w;e_{\ii},x)
\label{mubar}
\ee
\be
|\nb|=\sum_{\sigma}\int_{-\infty}^0\rmd\w\ D_{\sigma}^0(\w;e_{\ii},x)
\label{nbar}
\ee
\reseteqn
in terms of the MF spectral density $D^{0}_{\sigma}(\omega) =
-\frac{1}{\pi}\mbox{sgn}(\w)\mbox{Im}{\cal{G}}_{\sigma}(\omega)$. The class of diagrams retained
in practice for $\Sigma_{\sigma}(\omega) = \Sigma_{\sigma}[\{ {\cal{G}}_{\sigma}
\}]$ remains unchanged from previous work, and captures the key spin-flip
dynamics inherent to both the GFL and LM phases. For a full discussion of the
diagrams and their physical content the reader is referred to [15,25,27]. Shown in figure 2, they translate to
\begin{figure}
\begin{center}
\epsfig{file=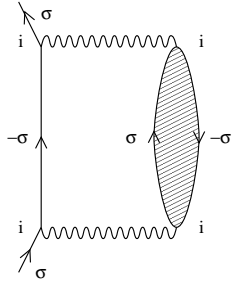,width=3cm}
\caption{Principal contribution to the dynamical $\Sigma_{\sigma}(\omega)$. Same notation as in figure 1; the transverse spin polarization propagator (see text) is shown as hatched.}

\label{diagrams}
\end{center}
\end{figure}
\be
\Sigma_{\sigma}(\w)=U^2\int_{-\infty}^{\infty}\frac{\rmd\w_1}{2\pi\im}\ \scrG_{-\sigma}(\w-\w_1)\Pi^{-\sigma\sigma}(\w_1).
\label{se}
\ee
Here $\Pi^{\sigma -\sigma}(\omega)$ is the transverse spin polarization propagator
given by an RPA-like particle-hole ladder sum in the transverse spin channel,
\be
\Pi^{\sigma-\sigma}(\w)=\ ^0\Pi^{\sigma-\sigma}(\w)[1-U ^0\Pi^{\sigma-\sigma}(\w)]^{-1}
\label{fullpi}
\ee
where $^{0}\Pi^{\sigma -\sigma}(\omega)$ is the bare particle-hole bubble, itself
expressed in terms of the broken symmetry MF propagators $\{ {\cal{G}}_{\sigma} \}$
and satisfying the Hilbert transform
\be
 ^0\Pi^{+-}(\w)=\int_{-\infty}^{\infty}\frac{\rmd\w_1}{\pi}\ \frac{\mbox{Im} ^0\Pi^{+-}(\w_1)\mbox{sgn}(\w_1)}{\w_1-\w^+}.
\label{pi0ht}
\ee
We add futher that since [15,27] $\Pi^{+-}(\omega) = \Pi^{-+}(-\omega)$ (and
likewise for the $^{0}\Pi^{\sigma -\sigma}(\omega)$), only
$\Pi^{+-}(\omega)$ need be considered henceforth.

  The functional dependence of the $\{ \tilde{\Sigma}_{\sigma}(\omega) \}$ on the
MF propagators, for both the GFL and LM phases, remains as detailed in [15] for
the symmetric model. For the GFL phase in particular, $\Pi^{+-}(\omega)$ obeys the
same Hilbert transform equation (4.13) as $^{0}\Pi^{+-}$ and equation (4.11) may thus be written
as
\be
\fl \Sigma_{\sigma}(\w)=U^2\int_{-\infty}^{\infty}\frac{\rmd \w_1}{\pi}\ \mbox{Im}\Pi^{+-}(\sigma\w_1)\left[\theta(\w_1)\scrG_{-\sigma}^-(\w_1+\w)+\theta(-\w_1)\scrG_{-\sigma}^+(\w_1+\w)\right]
\label{sigofw}
\ee
where
\be
\scrG_{\sigma}^{\pm}(\w)=\int_{-\infty}^{\infty}\rmd\w_1\ \frac{D^0_{\sigma}(\w_1)\theta(\pm\w_1)}{\w-\w_1\pm \im 0^+}
\label{gpmht}
\ee
denote the one-sided Hilbert transforms such that
${\cal{G}}_{\sigma}(\omega) = {\cal{G}}_{\sigma}^{+}(\omega) +
{\cal{G}}_{\sigma}^{-}(\omega)$.

\subsubsection*{Stability}
  The Hilbert transform equation (4.13) for $\Pi^{+-}(\omega)$  implies the stability condition $\pi$Re$\Pi^{+-}(\omega=0) =
\int^{\infty}_{-\infty}\rmd\omega_{1}$Im$\Pi^{+-}(\omega_{1})/|\omega_{1}|
>0$. From equation (4.12), together with Im$^{0}\Pi^{+-}(\omega =0) =0$, it follows
that this positivity is guaranteed only if $0<U$Re$^{0}\Pi^{+-}(\omega =0) \leq 1$.
An explicit expression for Re$^{0}\Pi^{+-}(0)$ is however readily obtained [15,27],
\be
U\mbox{Re}^0\Pi^{+-}(\w=0)=\frac{|\mb(e_{\ii},x)|}{|\mu|}
\label{stab2}
\ee
with $|\bar{\mu}(e_{\ii},x)|$ given by equation (4.10a). The pure MF local moment,
denoted by $|\mu_{0}| \equiv |\mu_{0}(e_{\ii})|$, is itself given as the self-consistent
solution of $|\mu| = |\bar{\mu}(e_{\ii},x=\frac{1}{2}U|\mu|)|$. Equation (4.16) thus
shows that the requisite stability is preserved for $|\mu| \geq |\mu_{0}|$,
and that the boundary to stability (when $U$Re$\Pi^{+-}(0;e_{\ii},x) =1)$ is precisely
the pure MF local moment $|\mu_{0}|$.

  When $|\mu| = |\mu_{0}(e_{\ii})|$ it follows from equation (4.16) that Im$\Pi^{+-}(\omega)$
(equation (4.12)) contains a zero-frequency pole. This is of course the essential signature
of the doubly degenerate LM phase, considered explicitly in the following section.
In the GFL phase by contrast, $|\mu| = |\mu_{0}| + \delta|\mu|$ is determined by
symmetry restoration (step (i) of \S 4.1.1); with $\delta|\mu| >0$ correctly found in
practice, as required for stability. In consequence Im$\Pi^{+-}(\omega)$ contains
a sharp resonance occurring at a non-zero frequency $\omega =\wm (\equiv \w_{*})$, the
low-energy Kondo scale for the GFL phase. This is all entirely in keeping with the general picture
presented hitherto for the symmetric PAIM, see e.g. figure 4 of [15]. As $U$ is increased
for any fixed asymmetry $\eta$ in the GFL phase, $\wm$ progressively
diminishes ($\delta|\mu| \rightarrow 0$) and vanishes at a critical
$U=U_{\cc}(r,\eta)$, where the resonance becomes an isolated pole at $\omega =0$ ($\delta|\mu| =0$). This signals the transition to the LM phase, to which we
now turn.

\subsection{LM phase}
Throughout the LM phase $|\mu| = |\mu_{0}(e_{\ii})|$ and hence $U$Re$\Pi^{+-}(0)=1$.
In consequence $\Pi^{+-}(\omega)$ contains an $\omega =0$ spin-flip pole
(of weight $Q$), together with a continuum contribution $^{\s}\Pi^{+-}(\omega)$ [15],
specifically
\be
\Pi^{+-}(\w)=-\frac{Q}{\w+\mbox{i}0^+}+\ ^{\mbox{\ssz{S}}}\Pi^{+-}(\w)
\label{5.24a}
\ee
with $Q = [U^{2}(\partial$Re${}^{0}\Pi^{+-}(\omega)/\partial\omega)_{\omega=0}
]^{-1} >0$. Im$^{\s}\Pi^{+-}(\omega)$ is given by equation (4.12) with
$\Pi^{+-} \rightarrow {}^{\s}\Pi^{+-}$, and the real/imaginary parts of $^{\mbox{\ssz{S}}}\Pi^{+-}(\w)$ are again
related by the Hilbert transform equation (4.13). The basic form for
$\Sigma_{\sigma}(\omega)$ in the LM phase then follows using equation (4.17) in
equation (4.11), namely
\be
\Sigma_{\sigma}(\w)=QU^2{\cal G}^{-\sigma}_{-\sigma}(\w)+^{\mbox{\ssz{S}}}\!\Sigma_{\sigma}(\w).
\label{LMse}
\ee
where $^{\s}\Sigma_{\sigma}(\omega)$ is given explicitly by equation (4.14) with
$\Pi^{+-} \rightarrow {}^{\s}\Pi^{+-}$.

  For any given $U$ and $\eta$ ($=1+2\epsilon_{\ii}/U$) in the LM phase
$U>U_{\cc}(r,\eta)$, $e_{\ii}$
is determined by requiring that
the change in number of electrons due to addition of the impurity
($\nimp$, equation (2.6)) is integral, as for the GFL phase discussed in \S 2.1.
Here however we require that $\nimp =1$, i.e.
\begin{equation}
\nimp = \frac{2}{\pi}\mbox{Im}\int^{0}_{-\infty}\rmd\omega \ G(\omega)
\left[1 - \frac{\partial\Delta(\omega)}{\partial\omega}\right] ~~~ =1
\end{equation}
in common with the local moment regime of the atomic
limit ($\epsilon_{\ii} <0$ and $\epsilon_{\ii} +U > 0$) to which we expect the LM
phase generally to be connected perturbatively in $V$. Satisfying equation (4.19) within
the approach presented thus determines $e_{\ii}$ in a manner directly analogous to
use of the Luttinger theorem to determine $e_{\ii}$ for the GFL phase as described
in \S 4.1.1 (and in fact $\nimp=1$ is found in practice to be the only possible
integer solution for $\nimp$ in the LM phase).

Symmetry is not of course restored in the doubly degenerate LM phase, i.e.\ $\st^{\subr}_{\up}(\w=0)\neq\st^{\subr}_{\down}(\w=0)$.  The difference  $\st^{\subr}_{\up}(0)-\st^{\subr}_{\down}(0)$ is however found to be sign definite for all $U>U_{\cc}$; such that as $U$ is decreased towards $U_{\cc}$, $|\st^{\subr}_{\up}(0)-\st^{\subr}_{\down}(0)|$ progressively diminishes and vanishes as the transition is approached, $U\ra U_{\cc}+$.  In this way the phase boundary between LM and GFL phases may be established, now coming from the LM side of the transition $U>U_{\cc}$.  And the resultant $U_{\cc}\equiv U_{\cc}(r,\eta)$ is correctly found to coincide precisely with that obtained coming from the GFL side, $U<U_{\cc}$, as the limit of solution of the symmetry restoration condition equation (4.5) for the GFL phase (as indeed is to be expected since as $U\ra U_{\cc}+$, the resultant vanishing of  $|\st^{\subr}_{\up}(0)-\st^{\subr}_{\down}(0)|$ corresponds to the `onset' of symmetry restoration).

\section{Scales, exponents and phase boundaries}
We turn now to results arising from the LMA specified above. Single-particle
dynamics will be analysed explicitly in \S  6ff. Here the behaviour of the relevant low-energy scales in the vicinity of the quantum critical point is first
considered, followed in \S 5.2 by resultant phase diagrams in the $(U,\eta)$
(or equivalently $(U,\epsilon_{\ii}))$ plane.

\subsection{Scales}
Our aim here is simply to establish the critical behaviour of the relevant
low-energy scales, specifically $\omega_{*}$ (in terms of which single-particle
dynamics exhibit universal scaling (\S s 3,6,7)); and, relatedly, the behaviour of the
renormalised level $\epsilon_{\ii}^{*} = \epsilon_{\ii} + \Sigma^{\subr}(0)$
(equation (2.14)). Unless stated to the contrary, all results given are representative
of the full range $0<r<1$ considered in the paper.
\begin{figure}
\begin{center}
\psfrag{ulabel}[bc][bc]{\hspace{0.0cm}\bf ${U}$}
\epsfig{file=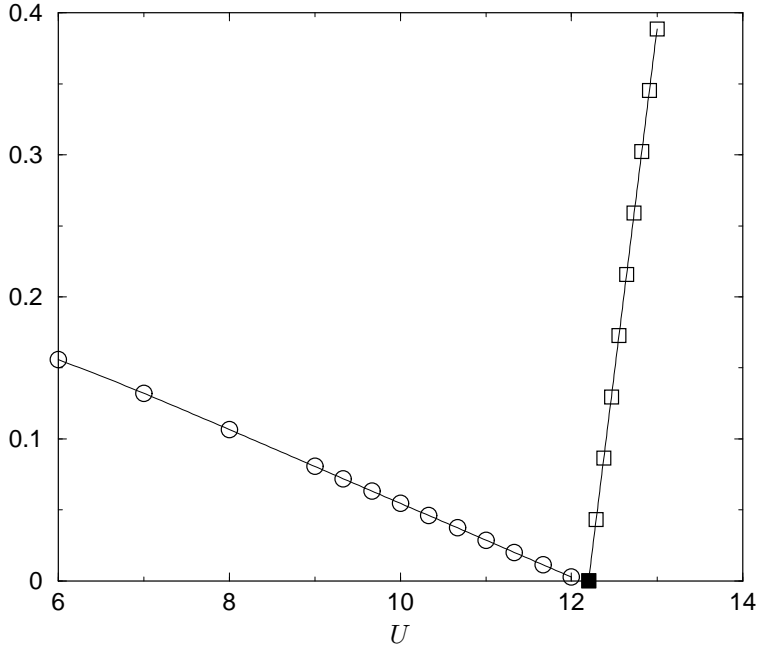,width=10cm}
\caption{Critical behaviour of the low-energy scale $\omega_{*}$ in the
vicinity of the QCP; illustrated for $r=0.3$ and asymmetry $\eta =0.2$ (critical $U_{\cc} \simeq 12.2$, solid square). $\omega_{*}^{r}$ \it vs \rm $U$ is shown:
for $U < U_{\cc}$ (GFL phase, $\omega_{*} \equiv \omega_{\m}$), $[\omega_{\m}/U]^{r}$
\it vs \rm $U$ is plotted (open circles); for $U>U_{\cc}$
(LM phase, $\omega_{*} \equiv \wl$),
$|\tilde{\Sigma}^{\subr}_{\uparrow}(0)-\tilde{\Sigma}^{\subr}_{\downarrow}(0)|
\propto \omega_{*}^{r}$ is plotted (open squares). In either case, $\omega_{*} \sim |U-U_{\cc}|^{\frac{1}{r}}$ as the transition is approached. See text for further details.}

\end{center}
\end{figure}

  We begin by considering the GFL phase, $U<U_{\cc}$. The critical behaviour of the
low-energy Kondo scale $\omega_{\m} \equiv \omega_{*}$ is illustrated in figure 3
for $r=0.3$ and an asymmetry $\eta = 1 -2|\epsilon_{\ii}|/U = 0.2$; the critical
$U_{\cc}(r,\eta) \simeq 12.2$ being marked as a solid square. Specifically,
$(\omega_{\m}/U)^{r}$ \it vs \rm $U$ (open circles) is shown, and seen to vanish
linearly
as $U \rightarrow U_{\cc}-$. The critical exponent $a$ of \S 3 ($\omega_{*} \sim u^{a}$
with $u = |1-U/U_{\cc}|)$ is thus $a = 1/r$, which agrees with the
NRG calculations of [18] (the result is also derivable
analytically for $r \ll 1$, see \S 7.1). We find also that
$\omega_{*} \equiv \omega_{\m}$ and the quasiparticle weight $Z$ are indeed related by
equation (3.8), namely that $Z \sim \omega_{*}^{1-r} \sim u^{(1-r)/r}$ as $u \rightarrow 0$.

  As argued in \S 3, the renormalised level $\epsilon_{\ii}^{*}$
should likewise vanish as the transition is approached, from
either phase. When considering the LM phase in particular it will
in general prove
helpful to define spin-dependent renormalised levels via
\begin{equation}
\epsilon_{\ii\sigma}^{*} = \epsilon_{\ii} + \tilde{\Sigma}^{\subr}_{\sigma}(\omega =0).
\end{equation}
The two-self-energies $\{ \tilde{\Sigma}_{\sigma}^{\subr}(\omega) \}$ are related to
the conventional single self-energy $\Sigma(\omega)$ by equation (2.11), from which
the $\{ \epsilon_{\ii\sigma}^{*} \}$ and the renormalised level
$\epsilon_{\ii}^{*} = \epsilon_{\ii} + \Sigma^{\subr}(0)$ are related by
\begin{equation}
\epsilon_{\ii}^{*} = \frac{\epsilon_{\ii\uparrow}^{*}\epsilon_{\ii\downarrow}^{*}}
{\frac{1}{2}[\epsilon_{\ii\uparrow}^{*}+\epsilon_{\ii\downarrow}^{*}]}.
\end{equation}
In the GFL phase (where symmetry is restored, $\tilde{\Sigma}^{\subr}_{\uparrow}(0)
=\tilde{\Sigma}^{\subr}_{\downarrow}(0)$) the renormalised levels all coincide,
$\epsilon_{\ii}^{*} = \epsilon_{\ii\sigma}^{*}$. For the LM phase by contrast,
$\epsilon_{\ii}^{*}, \epsilon_{\ii\uparrow}^{*}$ and $\epsilon_{\ii\downarrow}^{*}$
are in general distinct.

\begin{figure}
\begin{center}
\psfrag{xaxis}[bc][bc]{\large $U$}
\psfrag{yaxis}[bc][bc]{\large $\ $}
\psfrag{eidn}[b][b]{\hspace{0.3cm}${\epsilon_{\ii\dn}^*}$}
\psfrag{eiup}[b][b]{\hspace{0.3cm}${\epsilon_{\ii\up}^*}$}
\psfrag{ei}[b][b]{\hspace{0.5cm}${\ei^*}$}
\psfrag{ei1000}[b][b]{\hspace{0.9cm}${\ei^*}\times 10^3$}

\epsfig{file=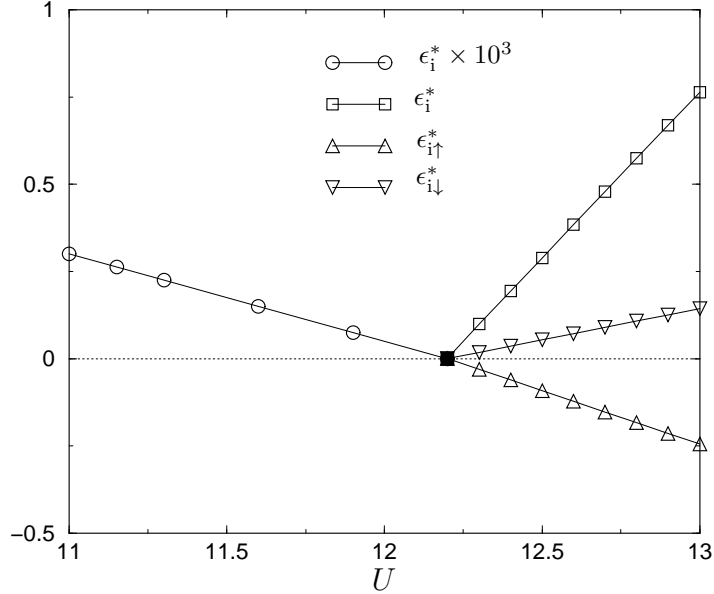,width=10cm}
\caption{Renormalised level $\epsilon_{\ii}^{*} = \epsilon_{\ii} + \Sigma^{\subr}(0)$
\it vs \rm $U$ in the vicinity of the QCP for $r=0.3$ and $\eta = 0.2$. GFL phase
points ($U <U_{\cc}$) are shown as circles, LM phase as squares. In the LM phase
the $\uparrow/\downarrow$-spin renormalised levels $\epsilon_{\ii\sigma}^{*} =
\epsilon_{\ii} + \tilde{\Sigma}^{\subr}_{\sigma}(0)$ are also shown, as up/down triangles.  All vanish linearly as the transition is approached.}
\label{}
\end{center}
\end{figure}

 Figure 4, again for $r=0.3$ and $\eta =0.2$,  shows the resultant $U$-dependence
of $\epsilon_{\ii}^{*}$ (open circles for $U < U_{\cc}$ in the GFL phase, squares for
$U>U_{\cc}$). The renormalised level is indeed seen to vanish as the transition
is approached from either phase, with an exponent $a_{1} =1$ ($|\epsilon_{\ii}^{*}|
\sim u^{a_{1}}$), in agreement with the inferences drawn in \S 3. For the GFL
phase in particular, $\omega_{*} \equiv \omega_{\m}$ and $|\epsilon_{\ii}^{*}|$
are thus related by equation (3.11), $|\epsilon_{\ii}^{*}| = k \omega_{*}^{r}$ (and where
the constant $k \equiv k(\eta)$ is an odd function of the asymmetry $\eta$,
such that $\epsilon_{\ii}^{*} =0$ by symmetry (\S 3) for the GFL phase of the
particle-hole symmetric
PAIM, $\eta =0$). For the LM phase $U > U_{\cc}$, figure 4 also shows the
spin-dependent renormalised levels $\epsilon_{\ii\sigma}^{*}$ (triangles); each
being seen to be sign definite and to vanish linearly as $U \rightarrow U_{\cc}+$.

  For the GFL phase, the low-energy (Kondo) scale $\omega_{*} \equiv \omega_{\m}$
arises as a direct consequence of symmetry restoration. The reader will however
notice that the corresponding scale $\omega_{*}$ for the LM phase, in terms of
which single-particle dynamics exhibit scaling and denoted
by $\omega_{*} \equiv \wl$, has not yet been specified. Here we simply
assert, and consider further in \S s 6ff, that $\wl$ is related to the
spin-dependent renormalised levels by
\begin{equation}
\omega_{*} \equiv \wl \propto |\epsilon_{\ii\uparrow}^{*} -
\epsilon_{\ii\downarrow}^{*}|^{\frac{1}{r}} = |\tilde{\Sigma}^{\subr}_{\uparrow}(0) -
\tilde{\Sigma}^{\subr}_{\downarrow}(0)|^{\frac{1}{r}}
\end{equation}
(or equivalently by $\wl \propto |\epsilon_{\ii\sigma}^{*}|^{\frac{1}{r}}$;
which equivalence follows since each $\epsilon_{\ii\sigma}^{*}$ vanishes linearly
as $U \rightarrow U_{\cc}+$, figure 4, and low-energy scales are equivalent up
to an arbitrary constant. The specific proportionality employed in practice
will also be specified in \S 6.2). Figure 3 also shows
$|\tilde{\Sigma}_{\uparrow}^{\subr}(0)-\tilde{\Sigma}_{\downarrow}^{\subr}(0)| \propto
\wl^{r}$ for $U > U_{\cc}$ (open squares); the overall figure thus showing the $U$-dependence
of $\omega_{*}^{r}$ in \it both \rm phases.

\subsection{Phase diagrams}

 We now consider representative phase diagrams arising from the LMA, obtained
as described in \S 4. As usual the range $0<r<1$ is considered; and we show in
particular that the essential characteristics of the phase diagrams divide into two
classes, namely $r<\frac{1}{2}$ and $r > \frac{1}{2}$.
\begin{figure}
\begin{center}
\psfrag{xaxis}[bc][bc]{\large $\ei$}
\psfrag{yaxis}[bc][bc]{\large $U$}
\psfrag{lm}[bc][bc]{\Large LM}
\psfrag{gfl}[bc][bc]{\Large GFL}
\psfrag{sc}[bc][bc]{\Large GFL}
\psfrag{ni0}[bc][bc]{\normalsize ($\nimp=0$)}
\psfrag{ni1}[bc][bc]{\normalsize ($\nimp=1$)}
\psfrag{ni2}[bc][bc]{\normalsize ($\nimp=2$)}
\epsfig{file=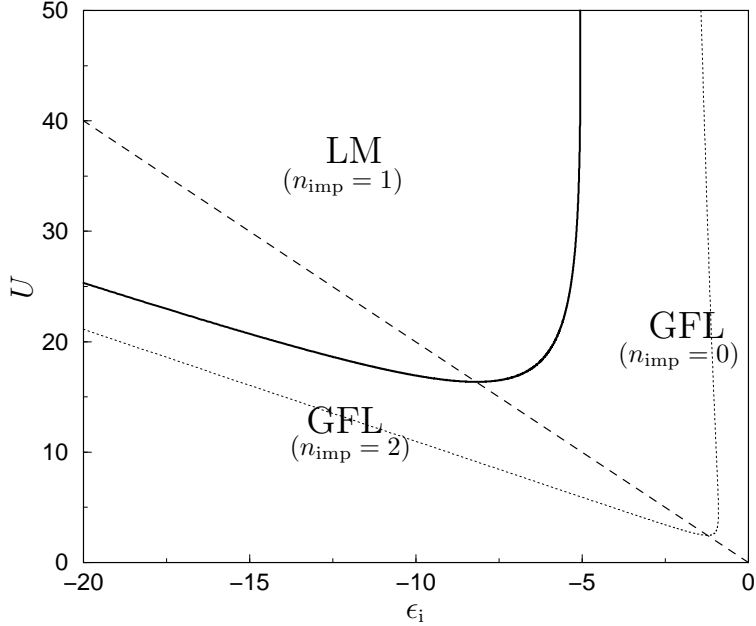,width=10cm}
\caption{LMA phase boundary in the $(U,\epsilon_{\ii})$-plane for a fixed
$r=0.2$ (solid line), representative of the general behaviour for $r<\frac{1}{2}$.
The dashed line marks the particle-hole symmetric case: $U=-2\epsilon_{\ii}$. The
corresponding MF phase boundary is shown as a dotted line.}

\label{pb1}
\end{center}
\end{figure}

  Figure 5 shows behaviour typical of $r<\frac{1}{2}$, the phase
boundary in the $(U,\epsilon_{\ii})$-plane for a fixed $r=0.2$.  As $\epsilon_{\ii}$
is progressively increased, the critical  $U_{\cc} \equiv U_{\cc}(r,\epsilon_{\ii})$
(solid line) first decreases, reaches a minimum at the particle-hole symmetric
line $\epsilon_{\ii} = -\frac{1}{2}U$ (shown as a dashed line), before rising sharply at
$\epsilon_{\ii} \simeq -5$ and eventually turning back on itself. This has two
interesting consequences. First that the LM phase is inaccessible by increasing
$U$ for any fixed $\epsilon_{\ii} \gtrsim -5$ in the present example. Second, for
fixed $\epsilon_{\ii}$ and due to the above `turnback', there is at large
$U$ a re-entrant transition back to the GFL phase (itself seen most clearly
in figure 18 below, where a low-$r$ phase boundary determined analytically
in \S 7.1 is shown). Both observations are in general accord with
results of recent NRG calculations [36]; and indeed are seen even at the pure
MF level of unrestricted Hartree-Fock, shown in figure 5 as a dotted line (although
the LM phase is strongly exaggerated by pure MF, the many qualitative deficiencies
of which are discussed e.g. in [15,25,27]).

\begin{figure}
\begin{center}
\psfrag{yaxis}[bc][bc]{\large $(8/\pi r)~{U}^{r-1}$}
\psfrag{xaxis}[bc][bc]{\large $\eta$}
\psfrag{LM}[bc][bc]{\Large LM}
\psfrag{GFL}[bc][bc]{\Large GFL}
\epsfig{file=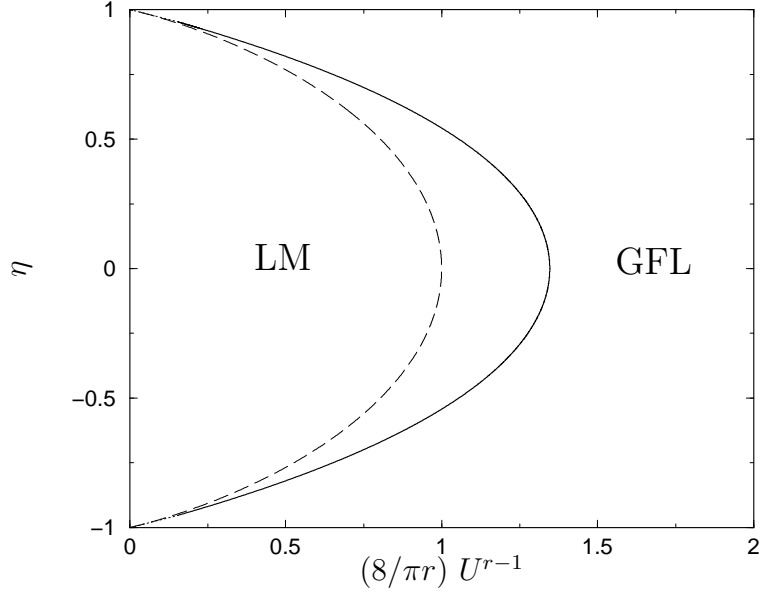,width=10cm}

\caption{Phase diagram in the $(U,\eta)$-plane for $r=0.2$:
$\eta$ \it vs \rm $(8/\pi r)U^{r-1}$ is shown (solid line).  The corresponding $r \ra 0$ result, obtained analytically in \S 7.1, is also shown (dashed line); see \S 7.1 for full discussion.}

\label{}
\end{center}
\end{figure}

  A rather more revealing portrayal of the phase boundary for fixed $r$ is shown in
figure 6 where, for $r=0.2$,  $U_{\cc} \equiv U_{\cc}(r,\eta)$ is displayed as a function
of the asymmetry $\eta$. Specifically, $\eta$ \it vs \rm
$(8/\pi r) U_{\cc}^{r-1}(r,\eta)$ is shown; the constant $8/\pi r$ simply serving
as a convenient `normalisation', it being known from previous LMA work [15]
that as $r \rightarrow 0$, $U_{\cc}^{r-1}(r,0) \sim \pi r/8$ for the symmetric PAIM
$\eta =0$ (which result is exact [15,16]). In contrast to $U_{\cc}(r,\epsilon_{\ii})$ shown
in figure 5, which is not single-valued in $\epsilon_{\ii}$ (leading to re-entrance
as above), figure 6 shows $U_{\cc}(r,\eta)$ to be a single-valued function of
the asymmetry (which is found for all $r$); and that LM states are always accessible
upon increasing $U$ for any fixed asymmetry $|\eta| <1$. In \S 7.1 the $\eta$-dependence
of $U_{\cc}(r,\eta)$ arising within the LMA will be discussed further and obtained analytically in the small-$r$ limit (which result is also shown in figure 6).


\begin{figure}
\begin{center}
\psfrag{xaxis}[bc][bc]{\large ${\epsilon}_{\ii}$}
\psfrag{yaxis}[bc][bc]{\large ${U}$}
\psfrag{x1axis}[bc][bc]{\small $(8/\pi r){U}^{r-1}$}
\psfrag{y1axis}[bc][bc]{ $\eta$}
\psfrag{lm}[bc][bc]{\Large LM}
\psfrag{gfl}[bc][bc]{\Large GFL}
\psfrag{lms}[bc][bc]{\small LM}
\psfrag{gfls}[bc][bc]{\small GFL}
\epsfig{file=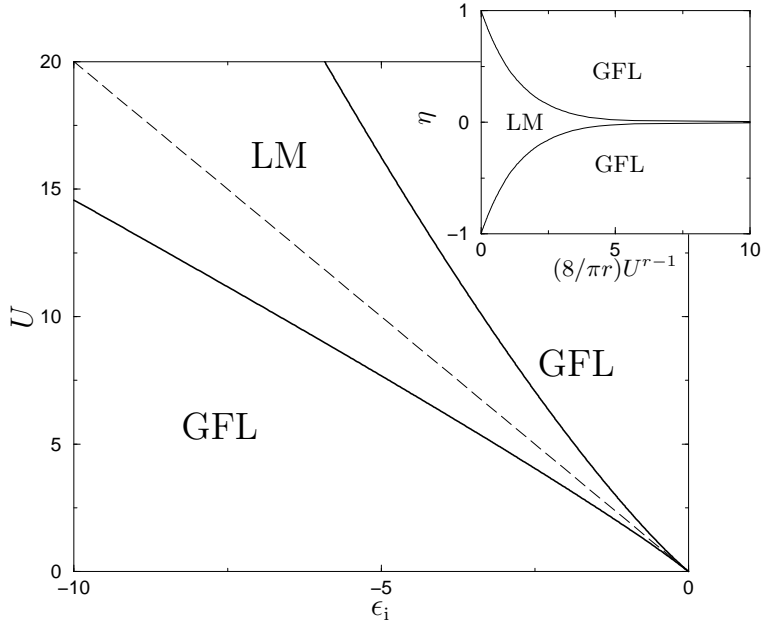,width=10cm}

\caption{Main figure: LMA phase boundary in the $(U,\epsilon_{\ii})$-plane
for $r=0.7$, respresentative of the general behaviour for $r>\frac{1}{2}$. The dashed
line marks the symmetric limit, $U=-2\epsilon_{\ii}$; in this case $(\eta =0)$ only
LM states are accessible. Inset: phase boundary in the $(U,\eta)$-plane,
$\eta$ \it vs \rm ($8/\pi r)U^{r-1}$ being shown.}
\label{pb2}
\end{center}
\end{figure}

Figure 6 also shows clearly that particle-hole symmetry is optimal for
LM formation --- $\eta =0$ corresponding to the lowest critical $U_{\cc}(r,\eta)$.
For the symmetric model it is however well known [10-20] that for $r>1/2$  the
GFL phase does not survive for any non-zero $U$. It is for this reason
that phase diagrams for $r\lessgtr 1/2$ are distinct, and
figure 7 illustrates the point. For $r=0.7$ the main figure (cf figure 5) shows the critical
$U_{\cc}(r,\epsilon_{\ii})$ \it vs \rm $\epsilon_{\ii}$; from which in particular the re-entrant behaviour upon increasing $U$ for fixed $\ei$ is again apparent.   And the
inset to figure 7 shows the $\eta$-dependence of the phase boundary (in direct parallel to
figure 6). The qualitative differences between the $r<1/2$ and $r>1/2$ phase
boundaries are self-evident.

\begin{figure}
\begin{center}
\psfrag{xaxis}[bc][bc]{\large $r$}
\psfrag{yaxis}[bc][bc]{\large ${U}^{-1}$}
\epsfig{file=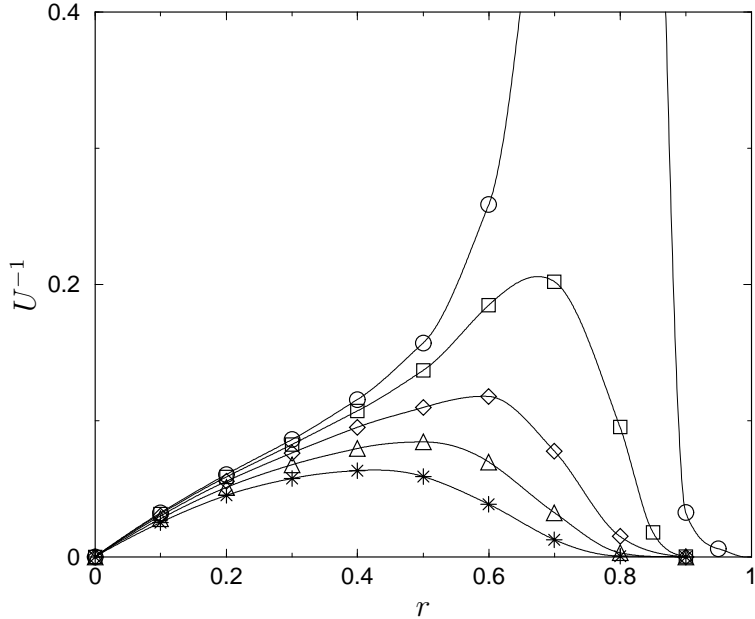,width=10cm}
\caption{$r$-dependence of the phase boundaries for fixed asymmetry
$\eta$: $U^{-1}$ \it vs \rm $r$ for (from top to bottom) $\eta = 0.1, 0.2, 0.3,
0.4$ and $0.5$. See text for details. Since the wide-band limit ($D=\infty$) is employed here, the largest $r$ considered in practice is $r=0.95$.}
\label{pb3}
\end{center}
\end{figure}

  The $r$-dependence of the phase boundaries
is illustrated in figure 8, which shows $U_{\cc}^{-1}(r,\eta)$ \it vs \rm
$r$ for a range of asymmetries $\eta$. For $\eta >0$ the resultant phase boundaries
have a maximum --- in qualitative accord with the NRG  results of
[13] for $U=\infty$ --- which moves to lower $r$ with increasing $\eta$. For the
symmetric case $\eta =0$ the present approach is known [15] to be deficient as
$r \rightarrow 1/2$, in not capturing the slow (logarithmic) divergence of
$U_{\cc}^{-1}(r\rightarrow \frac{1}{2} ,0)$ seen
in NRG calculations [12,13,16]; in turn reflecting the neglect of charge fluctuations
in the present LMA (which are necessarily important in this limit, since
$U_{\cc}(r\rightarrow \frac{1}{2},0) \rightarrow 0$). We do not however doubt at
least the qualitative validity of the results shown for $\eta >0$ where
$U_{\cc}(r,\eta)$ remains finite for all $r$ considered.

  Finally, note from figure 8 that for small $r$ the critical $U_{\cc}^{-1}(r,\eta)
\propto r$ for fixed asymmetry, with a coefficient that is
largest in the particle-hole symmetric case $\eta =0$. We will revisit the
matter analytically in \S 7.1, and show that this behaviour is intimately related
to the exponential asymptotics of the ubiquitously non-vanishing Kondo scale
characteristic of the $r=0$ metallic impurity model.

\seceq

\section{Single-particle dynamics}
 We turn now to LMA results arising for single-particle dynamics,
on all relevant energy scales and including in particular the many different
facets of their scaling behaviour as the underlying quantum phase transition is approached. The GFL and LM phases are considered separately in \S s 6.1 and 6.2 respectively, and dynamics precisely at the QCP in \S 6.3.

\subsection{GFL phase}
A typical GFL phase spectrum is given in figure 9: $\pi D(\omega)$ \it vs \rm
$\omega$, shown for $r=0.3$, an asymmetry $\eta = 1+2\epsilon_{\ii}/U =0.2$
and an interaction strength $U =11$ close to the critical
$U_{\cc}(r,\eta) \simeq 12.2$. Note first
that the change in number of electrons due to addition of the impurity is correctly
$\nimp =0$, as required by the Freidel sum rule equation (4.8) (where
$\epsilon_{\ii}^{*} = \epsilon_{\ii} + \tilde{\Sigma}^{\subr}_{\sigma}(0) >0$ for $\eta >0$).
By contrast, the local impurity charge $n = 2\int^{0}_{-\infty}\rmd\omega D(\omega)$
is $n \simeq 0.96$, close to the Kondo limit where $n \rightarrow 1$; showing in turn
(since $\nimp =0$) that essentially one electron is transferred from the conduction
band to the impurity upon its addition to the host.

\begin{figure}
\begin{center}
\psfrag{xaxis}[bc][bc]{{\large \bf $\w$}}
\psfrag{yaxis}[bc][bc]{{\large \bf ${\pi D(\w)}$}}
\epsfig{file=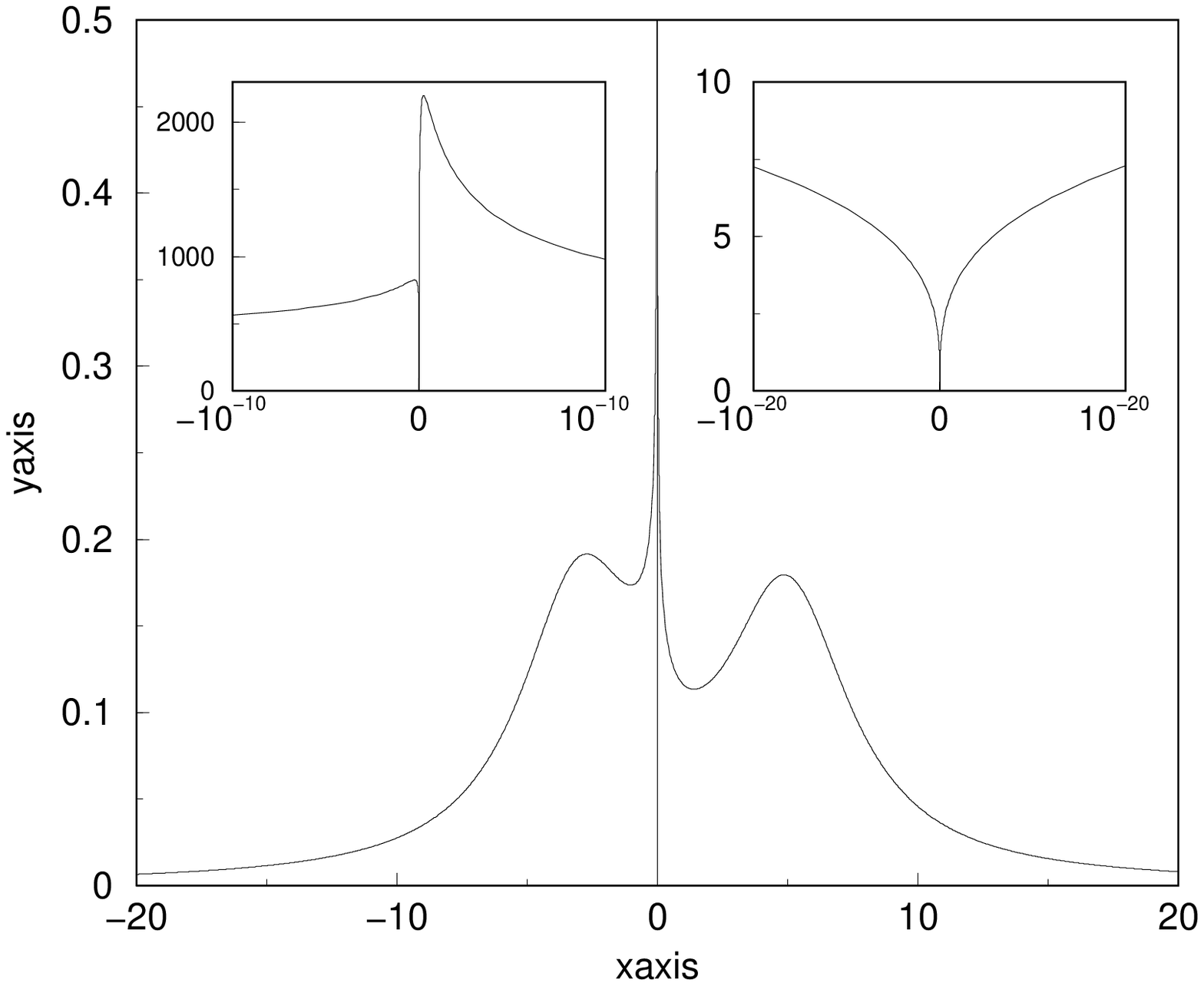,width=10cm}
\caption{GFL phase spectrum for $r=0.3$, asymmetry $\eta =0.2$
and interaction $U=11$ (the critical $U_{\cc}(r,\eta) \simeq 12.2$):
$\pi D(\omega)$ \it vs \rm $\omega$. Insets: spectral behaviour on the lowest
energy scales. See text for discussion.}
\label{gfl1}
\end{center}
\end{figure}
\begin{figure}
\begin{center}
\psfrag{xaxis}[bc][bc]{{\large \bf ${\w}$}}
\psfrag{yaxis}[bc][bc]{{\large \bf $\pi D(\w)$}}

\epsfig{file=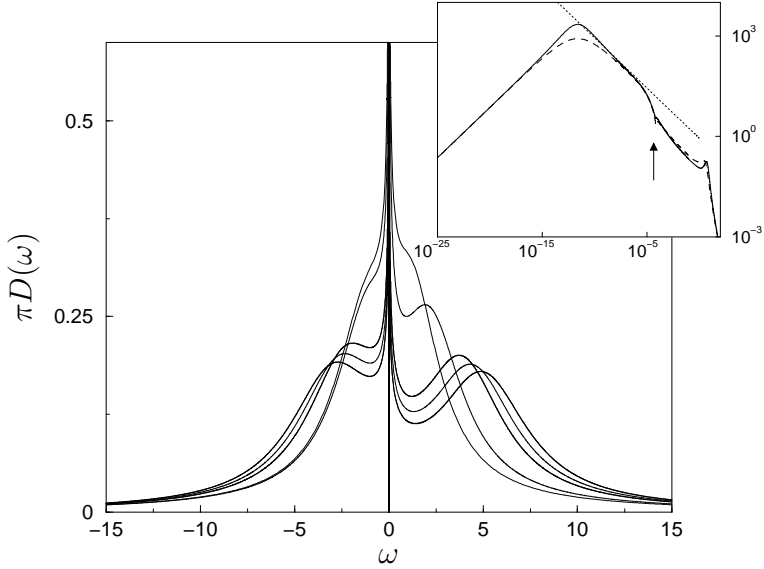,width=10cm}
\caption{GFL phase spectral evolution with $U$, for $r=0.3$ and
$\eta =0.2$ (the critical $U_{\cc} \simeq 12.2$): $\pi D(\omega)$ \it vs \rm
$\omega$ for $U = 11, 10, 9, 6$ and $5$ in obvious sequence.
Inset: $\pi D(\omega)$ \it vs \rm $|\omega|$ on a
logarithmic scale, for both $\omega >0$ (solid line) and $\omega <0$ (dashed).
$\w=\w_*$ ($\equiv \wm$) is indicated by a vertical arrow.  The function cos$^{2}(\frac{\pi}{2}r)|\omega|^{-r}$ is also shown (dotted line); see text for full discussion.}
\label{gfllog}
\end{center}
\end{figure}

On high energy scales the expected
Hubbard satellites are seen in the spectrum, centred on $\omega \simeq \epsilon_{\ii}$
and $\epsilon_{\ii} + U$; and correctly broadened two-fold over their counterparts
at pure MF level for the physical reasons discussed in [15,25]. Spectral evolution with
$U$ is illustrated in figure 10 where, for the same $r$ and $\eta$, the spectra are
shown upon decreasing $U$ further from the phase boundary into the GFL phase.
With diminishing interaction the Hubbard satellites become progressively less
pronounced as expected, and ultimately absorbed into the central resonance structure.
It is the latter that we now begin to consider in more detail, focusing first on the
lowest energy behaviour.

The left-hand inset of figure 9 shows the low-energy behaviour of the spectrum on
a greatly expanded scale. It consists of two pronounced peaks of unequal intensity,
one either side of the Fermi level $\omega =0$, with the more intense
peak arising for $\omega >0$. This is quite general, applying for all $\eta >0$
(where $\nimp =0$) and $r$, for $U$ close to $U_{\cc}(r,\eta)$.
For $\eta <0$ this situation is reversed: under $\eta \rightarrow -\eta$
as discussed in \S 2, the spectrum is simply reflected,
$D(\omega,\eta) = D(-\omega,-\eta)$ (and $\nimp =0 \rightarrow 2-\nimp =2$).
The greater peak then lies below, and the lesser above, the Fermi level.
For any asymmetry $\eta \neq 0$, where $\epsilon_{\ii}^{*} \neq 0$ in the GFL phase,
$D(\omega)$ has the ultimate low-$\omega$ behaviour $D(\omega) \propto |\omega|^{r}$;
given trivially (e.g. from equation (3.9)) by
\begin{equation}
\pi D(\omega) \stackrel{\omega \rightarrow 0}{\sim}
\frac{|\omega|^{r}}{{}\epsilon_{\ii}^{*}{}^{2}}
\end{equation}
and symmetric in $\omega$. This is illustrated in the right inset to figure 9
(and is of course in contrast to the symmetric PAIM [12,14-16,18,20] $\eta =0$ (and
$\epsilon_{\ii}^{*} =0$), for which
$D(\omega) \propto |\omega|^{-r}$ as $|\omega| \rightarrow 0$, equation (3.5)).

 The qualitative features of the preceding paragraph are in agreement with
results of recent NRG calculations [18]. We now show that they may be understood
using the simple low-$\omega$ quasiparticle form for the impurity
spectrum, discussed in \S 3,  that is characteristic of the GFL phase. Following
that however, we point out the severe inadequacies of the quasiparticle form
in understanding the GFL phase scaling spectrum on all relevant frequency
scales, and in particular its essential irrelevance in
understanding the behaviour of the spectrum at the quantum critical point
(which itself is considered in \S 's 6.3,7.2).

The quasiparticle spectrum is given in scaling form by equation  (3.9), as a function of
$\tilde{\omega} = \omega/\omega_{*}$ with $\omega_{*} = \wm$ the low-energy
Kondo spin-flip scale characteristic of the GFL phase (determined via
symmetry restoration as described in \S 4); and with $k_{1} = \omega_{*}^{1-r}/Z$
(equation  (3.8)) a constant of order unity as the transition is approached,
where $\omega_{*}$ and $Z$ separately vanish (\S 5).
The quasiparticle form is of course restricted by construction to
$|\tilde{\omega}| \ll 1$ (which regime is now considered); and for present
purposes equation  (3.9) may be replaced with negligible error (as shown below) by
\begin{equation}
\pi \omega_{*}^{r}D(\omega) \simeq \frac{|\tilde{\omega}|^{r}}
{[\frac{\epsilon_{\ii}^{*}}{\omega_{*}^{r}} -\mbox{sgn}(\omega)\beta(r)|\tilde{\omega}|^{r}]^{2}
+ |\tilde{\omega}|^{2r}}
\end{equation}
where $\beta(r) =$ tan$(\frac{\pi}{2}r)$ as usual (and with
$k =|\epsilon_{\ii}^{*}|/\omega_{*}^{r}$ likewise finite as $U \rightarrow
U_{\cc}(r,\eta)-$, as shown explicitly in \S 5). Equivalently and conveniently,
equation  (6.2) may be recast as
\begin{equation}
\pi |\epsilon_{i}^{*}| D(\omega) \simeq \frac{|\omega'|^{r}}
{[1 - \mbox{sgn}(\omega)\beta(r)|\omega'|^{r}]^{2} + |\omega'|^{2r}}
\end{equation}
with $\omega' = \omega/{}|\epsilon_{\ii}^{*}|{}^{\frac{1}{r}}$
(and where $\epsilon_{\ii}^{*} >0$ has been taken, as appropriate to $\eta >0$).
Equations (6.2) or (6.3) capture the low-energy `two-peak' spectral structure
illustrated in the insets to figure 9. Equation (6.1) is of course
trivially recovered as $|\omega| \rightarrow 0$; and from equation  (6.3) the resultant
spectrum has maxima at $\omega = \pm T^{*}$, where
\begin{equation}
T^{*} = [|\epsilon_{\ii}^{*}| cos(\mbox{$\frac{\pi}{2}$}r)]^{\frac{1}{r}}.
\end{equation}
For any non-zero asymmetry $\eta$, the low-energy scale $T^{*}$ is proportional
to $\omega_{*}$ and is thus equivalent to it
(as follows from equation  (3.11), noting that $k$ $(\equiv k(r,\eta))$ is odd in $\eta$
such that for the particle-hole symmetric model $\eta =0$, $\epsilon_{\ii}^{*}
=0$ as in \S 3). We have here introduced $T^{*}$ simply to make connection to the
NRG calculations of [18] for the asymmetric PAIM, where the low-energy scale was
\it defined \rm as the location of the spectral maxima.

\begin{figure}
\begin{center}
\psfrag{xaxis}[bc][bc]{{\large \bf $\w/T^*$}}
\psfrag{yaxis}[bc][bc]{{\large \bf $\pi|\ei^*| D(\w)$}}
\epsfig{file=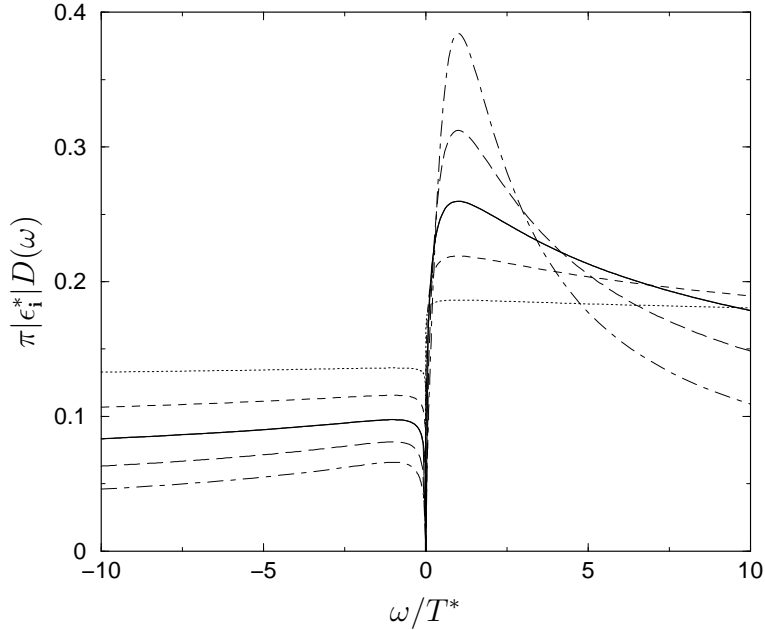,width=10cm}
\caption{
Low-frequency GFL scaling spectra:
$\pi |\epsilon_{\ii}^{*}|D(\omega)$
\it vs \rm $\omega/T^{*}$ (with the low-energy scale $T^{*}$ defined in equation
(6.4)), shown for $r=0.1$ (dotted line), $0.2$ (short dash),
$0.3$ (solid), $0.4$ (long dash) and $0.5$ (point dash).
For the case of $r=0.3$, explicit comparison is also
made to the approximate quasiparticle form equation (6.3); it is however
indistinguishable from the full LMA results (which were obtained explicitly for
$\eta =0.2$ but for $\eta \neq 0$ are in fact $\eta$-independent over the
$\omega/T^{*}$-range shown, see equation (6.3)).}
\label{prelimsc}
\end{center}
\end{figure}

 In figure 11, LMA results for $\pi |\epsilon_{\ii}^{*}| D(\omega)$
as the transition is approached are given, plotted \it vs \rm
$\omega/T^{*}$ for a range of different $r$'s (and noting that
in all cases the range $|\omega|/T^{*} < 10$ shown corresponds
to $|\omega|/\omega_{*} \ll 1$, since the constant
$k=|\epsilon_{\ii}^{*}|/\omega_{*}^{r} \sim (T^{*}/\omega_{*})^{r}$ is found
in practice to be small compared to unity). We emphasise that the figure shows
`full' LMA results; that the spectra are of course scaling spectra,
obtained as $U \rightarrow U_{\cc}(r,\eta)-$ and as such universally
dependent on $\omega/T^{*}$ (or equivalently $\tilde{\omega}$) alone;
and that the exponent $b$ of \S 3 (equation (3.2)) is indeed $b=r$.
As seen from figure 11 the asymmetrically
disposed spectral maxima are indeed peaked at $\omega/T^{*} = \pm 1$, and
for the particular case of $r=0.3$ the figure compares explicitly the LMA
results to the approximate quasiparticle form equation (6.3). This is not perceptibly
distinguishable from the full LMA results, showing that equation (6.3) is indeed accurate
(the same being true for the other $r$'s shown).

  For sufficiently low $\omega/T^{*}$ or $\tilde{\omega}$, the GFL phase scaling
spectrum is thus explicable in terms of the  simple approximate quasiparticle
form equations (6.2) or (6.3). An obvious question then arises: can this form be extrapolated beyond its strict confines, to
infer the leading \it large \rm  $\omega/T^{*}$ (or $\tilde{\omega}$) form
of the scaling spectrum  which --- as shown on general grounds in \S 3 (equations (3.2-4))
--- determines the behaviour of the spectrum precisely \it at \rm the quantum
critical point where the low-energy scale $\omega_{*} \propto T^{*} =0$, and
the generic form of which is given by equation (3.4)($\pi D(U_{\cc};\omega) =
C(r;\eta)|\omega|^{-r})$? The reader will notice for example that the leading high frequency
behaviour predicted from equations (6.2,3) is
$\pi \omega_{*}^{r} D(\omega) \sim$ cos$^{2}(\frac{\pi}{2}r)|\tilde{\omega}|^{-r}$;
from which the $\omega_{*}$ scale drops out of either side, resulting in an
apparent QCP spectrum that is indeed of the required general form equation (3.4),
but with a coefficient $C(r,\eta) =$cos$^{2}(\frac{\pi}{2}r)$.  As seen from equation (3.5) the resultant QCP spectrum, if correct,
would in fact coincide precisely with the {\it non-interacting} limit of the particle-hole symmetric ($\eta =0$) PAIM.

\begin{figure}
\begin{center}
\psfrag{yaxis}[bc][bc]{{\large \bf $\pi{\w_*}^r D(\w)$}}
\psfrag{xaxis}[bc][bc]{{\large \bf $\w/\w_*$}}
\epsfig{file=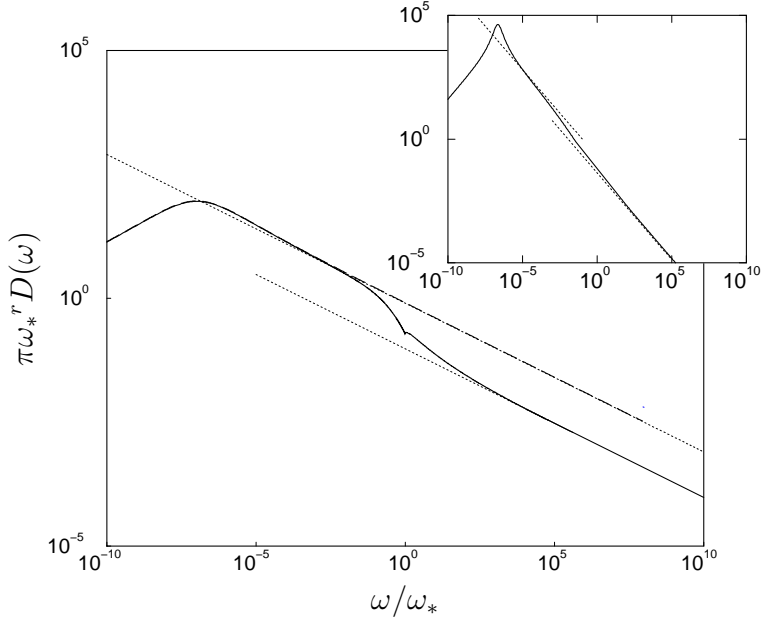,width=10cm}
\caption{Full GFL phase scaling spectrum for $r=0.3$ and $\eta =0.2$:
$\pi \omega_{*}^{r}D(\omega)$ \it vs \rm $\tilde{\omega}=\omega/\omega_{*}$ ($\w_*\equiv\wm$) plotted for $\tilde{\w}>0$
on a logarithmic scale (solid line) and encompassing all relevant $\tilde{\omega}$
regimes. The approximate quasiparticle form equation (6.2) is shown as a dashed
line; the upper dotted line shows the $\eta =0$ non-interacting behaviour
cos$^{2}(\frac{\pi}{2}r)|\tilde{\omega}|^{-r}$. For
$|\tilde{\omega}| \gg 1$ the ultimate high-frequency behaviour of the scaling
spectrum is also $\propto |\tilde{\omega}|^{-r}$ but with a different coefficient
$C(r,\eta)$ that is $r$- and $\eta$-dependent; the lower dotted line shows the latter
behaviour extrapolated back to lower $\tilde{\omega}$. Full discussion in text.  Inset: corresponding results for $r=0.7$ and $\eta=0.4$, same notation as main figure.}

\end{center}
\end{figure}

 The latter conclusions are not in fact correct, but the above discussion is
nonetheless informative as now shown. Figure 12 gives the full LMA scaling
spectrum for $r=0.3$ and
$\eta =0.2$: $\pi \omega_{*}^{r}D(\omega)$ for $\omega >0$ (solid line)
as a function of $\tilde{\omega} = \omega/\omega_{*}$, on a logarithmic scale
encompassing the entire relevant $\tilde{\omega}$ range (and with $\tilde{\omega}
=1$ seen as a small `dip' in the spectrum, a well known and harmless artefact of
the RPA-like form for $\Pi^{+-}$ employed here [15,25-27]). The figure also shows the
approximate quasiparticle form equation (6.2) (dashed line), together with its asymptotic
behaviour $\pi \omega_{*}^{r}D(\omega)
\sim$cos$^{2}(\frac{\pi}{2}r)|\tilde{\omega}|^{-r}$
mentioned above (upper dotted line). The full LMA scaling spectrum is well
captured by the quasiparticle form up to $\tilde{\omega} \simeq 10^{-1}$,
below which is indeed seen to exist an appreciable `intermediate' $\tilde{\omega}$-range over
which the behaviour $\pi \omega_{*}^{r}D(\omega)
\sim$cos$^{2}(\frac{\pi}{2}r)|\tilde{\omega}|^{-r}$ clearly arises. For larger
frequencies however, the scaling spectrum clearly departs from
$\sim$cos$^{2}(\frac{\pi}{2}r)|\tilde{\omega}|^{-r}$ behaviour; and for
$\tilde{\omega} \gg 1$ is seen to cross over into ultimate large-$\tilde{\omega}$
behaviour that is again $\propto |\tilde{\omega}|^{-r}$ --- following generally
from the fact that the exponent $b=r$ as explained in \S 3 --- but with a
prefactor $C(r,\eta)$ that differs from cos$^{2}(\frac{\pi}{2}r)$ and is
found in practice to be $\eta$-dependent.

  The above behaviour is not particular to the case $r=0.3$ illustrated, being
found to arise throughout the range $0<r<1$ considered (the inset to figure 12 shows corresponding results for $r=0.7$ and $\eta=0.4$). The asymmetric GFL phase
scaling spectrum thus contains two regimes of $\sim |\tilde{\omega}|^{-r}$ behaviour:
that at lower $\tilde{\omega}$ ($\ll 1$) corresponds to symmetric ($\eta =0$) non-interacting limit behaviour, while that arising for
$\tilde{\omega} \gg 1$ determines the QCP spectrum.  This `two $|\tilde{\omega}|^{-r}$ regime' behaviour is in fact known [20] to arise
also in the pure symmetric PAIM, $\eta =0$ (where the
cos$^{2}(\frac{\pi}{2}r)|\tilde{\omega}|^{-r}$ regime persists all the way down to
$\omega =0$, reflecting the exact low-$\omega$ behaviour equation (3.5) arising in that
case). Further consideration of the matter will be given in \S 6.3, and the
$(r,\eta)$-dependence of the QCP spectrum arising within the LMA  will be determined
analytically for small $r$ in \S 7.2.

Finally, we point out that figure 12 shows the scaling spectrum for
$\omega >0$, but that it is not of course symmetric in $\tilde{\omega}$ (or
$\omega/T^{*}$) as is obvious e.g.\ from figure 11. The sgn($\omega$)-dependence
of the GFL phase spectrum is illustrated further in the inset to figure 10
above (for $[r, \eta] = [0.3, 0.2]$ and $U=11$, quite close to $U_{\cc} \simeq 12.2$).
Here the spectrum $\pi D(\omega)$ \it vs \rm $|\omega|$ (i.e.\ unscaled)
is again shown on a log-scale, for both $\omega >0$ (solid
line) and $\omega <0$ (dashed). For each sgn($\omega$) the above `intermediate'
behaviour $\sim$cos$^{2}(\frac{\pi}{2}r)|\omega|^{-r}$ arises (shown as a dotted
line). Beyond that however, a crossover to $\sim C(r,\eta)|\omega|^{-r}$ behaviour
occurs; with the $C(r,\eta)$ dependent not only on $r$ and $\eta$ but
also for $\eta \neq 0$ on sgn($\omega$) (which behaviour we add is not precluded by the
general considerations of \S 3, and which is likewise considered further in the
following sections). Since the spectrum here is shown in unscaled form the
non-universal Hubbard satellites are also ultimately visible on the figure, likewise
asymmetrically disposed about $\omega =0$ because $\eta \neq 0$.   We also add here that while results from the present LMA
are in general agreement with the NRG calculations of [18], there is
one point on which we differ. That we find the coefficient
$C(r,\eta)$ to depend on $\eta$ (albeit fairly
weakly), is consistent with a line of ($\eta$-dependent) critical fixed
points. Numerical RG calculations for the pseudogap Kondo model [13] appear by
contrast to be consistent  for any $\eta >0$ with a single critical fixed
point, and hence in effect no $\eta$-dependence in $C(r,\eta)$.

\subsection{LM phase}
We turn now to the LM phase, considering first spectral evolution on `all scales'
as $U$ is progressively decreased for fixed asymmetry $\eta$, towards the
LM/GFL transition at
$U_{\cc}(r,\eta)$. This is illustrated in figure 13, again for $r=0.3$ and $\eta =0.2$
($U_{\cc} \simeq 12.2$), although the behaviour shown is representative of the full
$r$-range considered.
$\pi D(\omega)$ \it vs \rm $\omega$ is shown, for
$U=50, 20, 17$ and $14$.
\begin{figure}
\begin{center}
\psfrag{xaxis}[bc][bc]{{\large \bf ${\w}$}}
\psfrag{yaxis}[bc][bc]{{\large \bf $\pi D(\w)$}}
\psfrag{a}[bc][bc]{\large (a)}
\psfrag{b}[bc][bc]{\large (b)}
\psfrag{c}[bc][bc]{\large (c)}
\psfrag{d}[bc][bc]{\large (d)}

\epsfig{file=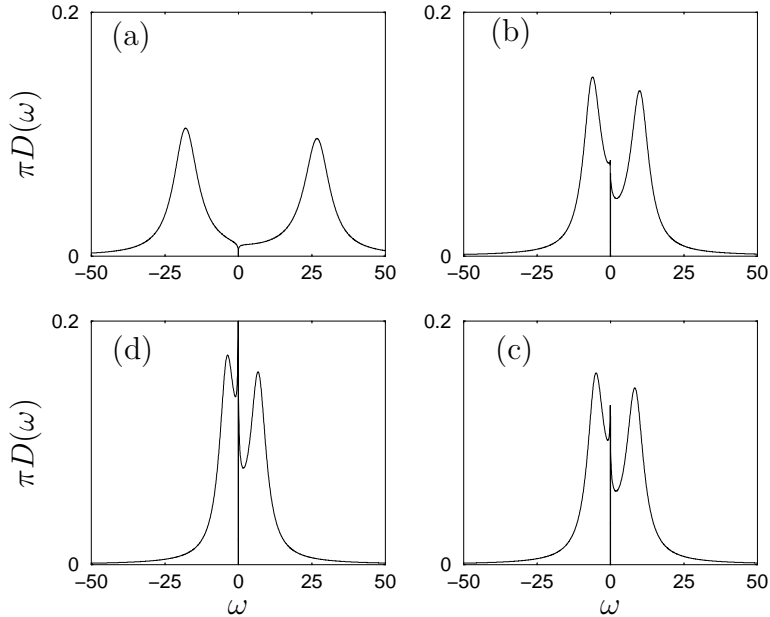,width=10cm}
\caption{LM phase spectral evolution with $U$, for $r=0.3$ and
$\eta =0.2$ ($U_{\cc} \simeq 12.2$): $\pi D(\omega)$ \it vs \rm $\omega$ for
$U=50,20,17,14$ (clockwise from top left).}
\label{}
\end{center}
\end{figure}
For the highest $U$ shown, deep in the LM phase, the well
developed Hubbard satellites dominate the spectrum entirely, although the ultimate
low-$\omega$ spectral behaviour vanishes as $\propto |\omega|^{r}$ throughout
the LM phase as noted in \S 3. With decreasing $U$ however, a low-energy spectral
structure is seen to develop in the vicinity of the Fermi level. It becomes increasingly
pronounced as $U \rightarrow U_{\cc}+$ and as shown below is the precursor, in the
LM phase, of the $|\omega|^{-r}$ divergent behaviour characteristic of the
QCP itself.

  For $U = 12.3$, close to $U_{\cc}$, $\pi D(\omega)$ \it vs \rm $\omega$ is shown
in figure 14; the left inset showing the low-energy part of the spectrum. The
latter is clearly redolent of its counterpart for the GFL phase (figure 9), the\begin{figure}
\begin{center}
\psfrag{xaxis}[bc][bc]{{\large \bf ${\w}$}}
\psfrag{yaxis}[bc][bc]{{\large \bf $\pi D(\w)$}}
\epsfig{file=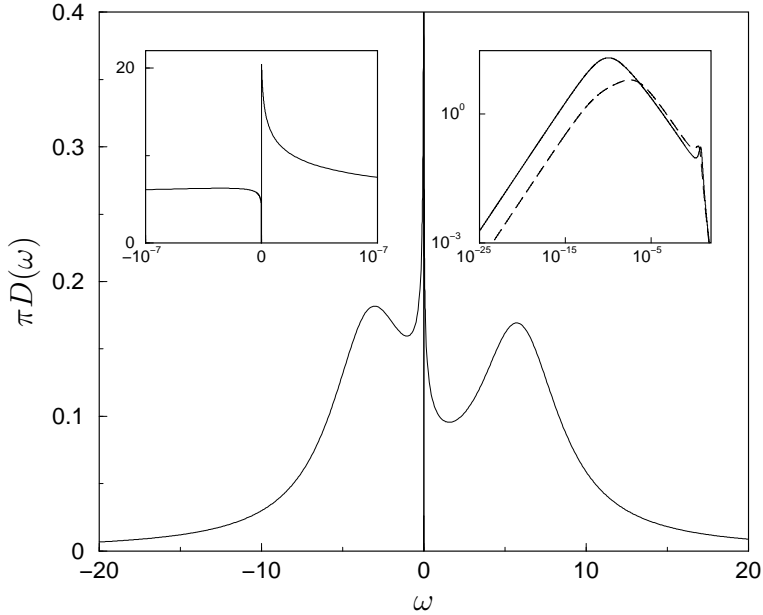,width=10cm}
\caption{LM phase spectrum $\pi D(\omega)$ \it vs \rm $\omega$ for
$r=0.3$, $\eta =0.2$ and interaction $U =12.3$ close to the critical $U_{\cc} \simeq
12.2$. Left inset: low-energy behaviour on an expanded scale. Right inset:
$\pi D(\omega)$ \it vs \rm $|\omega|$ on a logarithmic scale, for both
$\omega >0$ (solid line) and $\omega <0$ (dashed). Full discussion in text.}
\label{lmspec}
\end{center}
\end{figure}
low-energy behaviour again consisting of two peaks of unequal intensity, the more
intense peak arising for $\omega >0$ (and \it vice versa \rm for $\eta <0$).
In contrast to the GFL phase however, the positions of the peak maxima
are not symmetrically disposed about the Fermi level; reflecting
the fact that in the LM phase, where symmetry is not restored, the spin-dependent
renormalised levels $\epsilon_{\ii\sigma}^{*} = \epsilon_{\ii} +
\tilde{\Sigma}^{\subr}_{\sigma}(\omega =0)$ do not coincide (as shown e.g.\ in figure 4).
In addition, and again in contrast to the GFL phase, we find within the LMA that
while the lowest-$\omega$ spectral behaviour is $\propto |\omega|^{r}$ as noted above,
the coefficient thereof depends on sgn($\omega$). The origins of this are readily
seen using equations (2.9,10), which give the leading low-$\omega$ behaviour
\begin{equation}
{\pi D(\omega) \stackrel{\omega \rightarrow 0}{\sim}  \mbox{$\frac{1}{2}$}  }
\sum_{\sigma}\frac{[\Delta_{\subi}(\omega)+\Sigma^{\subi}_{\sigma}(\omega)]}
{[\epsilon_{\ii\sigma}^{*}]^{2}}
\end{equation}
with the LM phase $\Sigma_{\sigma}(\omega)$ given explicitly within the LMA
considered here by equation (4.18). It is the first term of the latter that controls
the low-$\omega$ behaviour of $\Sigma^{\subi}_{\sigma}(\omega)$ which, using equations
(4.6,15), is found to be of form $\Sigma^{\subi}_{\sigma}(\omega) \propto |\omega|^{r}$
with a coefficient dependent on sgn($\omega$) (as shown explicitly for $\eta =0$
in [15]); and which, via equation (6.5), generates corresponding behaviour in $D(\omega)$.

\begin{figure}
\begin{center}
\psfrag{xaxis}[bc][bc]{{\large \bf ${\w/\w_*}$}}
\psfrag{yaxis}[bc][bc]{{\large \bf $\pi \w_*^r D(\w)$}}
\epsfig{file=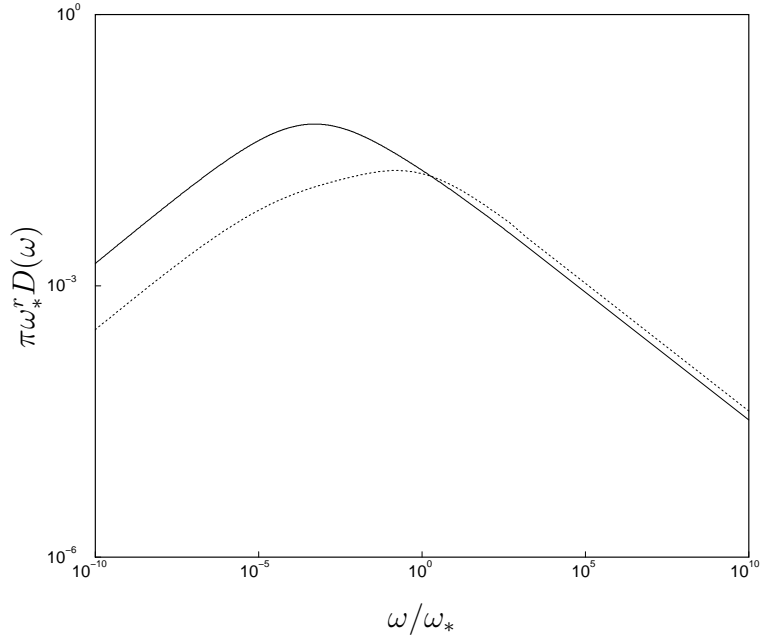,width=10cm}
\caption{Full LM phase scaling spectrum for $r=0.3$ and $\eta =0.2$,
obtained as $U \rightarrow U_{\cc}(r,\eta)+$: $\pi \omega_{*}^{r}D(\omega)$
\it vs \rm $\tilde{\omega} = \omega/\omega_{*}$ ($\w_*\equiv\wl$) on a logarithmic scale, for
both $\tilde{\omega} >0$ (solid line) and $\tilde{\omega} <0$ (dashed).}
\label{}
\end{center}
\end{figure}

 The low-$\omega$ peaks in figure 14 (left inset) mark the onset of
a crossover from the asymptotic $\propto |\omega|^{r}$ behaviour, to an $|\omega|^{-r}$
regime which persists essentially up to non-universal frequency scales on the order
$|\omega| \sim 1$ ($\equiv \Delta_{0}$). This is evident from the
right inset to figure 14 where the spectrum is plotted on a logarithmic scale:
the crossover from $D(\omega) \propto |\omega|^{r}$ to $\propto |\omega|^{-r}$
behaviour (likewise dependent on sgn($\omega$), albeit more weakly so) is clear,
as too are the non-universal Hubbard satellites on the highest energy scales.

  The spectra shown in figures 13, 14 are of course displayed on an `absolute'
scale, i.e. shown \it vs \rm $\omega$ in units of the hybridization strength
$\Delta_{0} \equiv 1$. But the LM phase is also characterised (\S 5) by a
vanishing low-energy scale $\omega_{*} \equiv \wl$ as the LM/GFL transition
is approached, so the corresponding LM phase scaling spectrum
is readily obtained in direct parallel to that for the GFL phase:
on decreasing $U$ progressively towards $U_{\cc}+$, $\omega_{*}$ steadily decreases,
but for fixed $\eta$ the spectra again exhibit universal scaling in terms of $\tilde{\omega} =
\omega/\wl$ alone (and with non-universal features such as the Hubbard satellites
thus `projected out' as usual). As anticipated throughout we find that it is indeed
$\pi \wl^{r} D(\omega)$ \it vs \rm $\tilde{\omega}$ which exhibits scaling
(whence, as for the GFL phase, the exponent $b$ of \S 3 is again found to be $b=r$
as required on general grounds); and with $\wl$ given by
the form equation (5.22). Specifically we take $\wl \equiv \omega_{*}$ to
be given by
\begin{equation}
\wl = |{\mbox{$\frac{\pi r}{8}$}}[\epsilon_{\ii\uparrow}^{*}-
\epsilon_{\ii\downarrow}^{*}]|^{\frac{1}{r}}
= |{\mbox{$\frac{\pi r}{8}$}}[\tilde{\Sigma}^{\subr}_{\uparrow}(0) -
\tilde{\Sigma}^{\subr}_{\downarrow}(0)]|^{\frac{1}{r}}
\end{equation}
where our (free) choice of prefactor, $\frac{\pi r}{8}$, is chosen to make
ready connection to the analytic results for small-$r$ obtained in \S 7.
The resulant LM phase scaling spectrum for $r=0.3, \eta =0.2$ is shown in figure 15,
showing clearly both the $\sim |\tilde{\omega}|^{-r}$ behaviour at low-$\tilde{\omega}$
and the crossover to the ultimate large-$\tilde{\omega}$ behaviour $\propto
|\tilde{\omega}|^{-r}$.

\subsection{Quantum critical point}
Which brings us to the quantum critical point itself. From the previous discussion
we would naturally expect GFL and LM phase spectra arbitrarily close to the QCP to
be indistinguishable for scales $|\omega| \gg \omega_{*}$. This is
evident in figure 16 which shows spectra $\pi D(\omega)$
\it vs \rm $\omega$ close to and on either side of the transition [$U_{\cc}(r =0.3,\eta
=0.2) \simeq 12.2$]; for $U =12.0$ in the GFL phase and $U=12.4$ for the LM phase.
On higher energy scales the spectra are indeed almost indistinguishable --- limited only
by the practical numerical inconvenience of dealing with the arbitrarily small but
strictly non-vanishing $\omega_{*}$ scales in the immediate vicinity of the transition.

  The inset to figure 16 shows the corresponding $\omega >0$ spectra on a logarithmic
scale. The leading low-$\omega$ behaviour $\pi D(\omega) \propto |\omega|^{r}$ is
clear, arising for each phase but with different coefficients for the GFL and LM
phases (and each such being $r$- and $\eta$-dependent). The figure also shows the
$\pi D(\omega) \propto |\omega|^{-r}$ behaviour arising for $\omega \gg \omega_{*}$
in either phase; and, most importantly, that the coefficient of this $|\omega|^{-r}$
behaviour is \it identical \rm for each phase --- as argued on general gounds in \S 3.

\begin{figure}
\begin{center}
\psfrag{xaxis}[bc][bc]{{\large \bf $\w$}}
\psfrag{yaxis}[bc][bc]{{\large \bf $\pi D(\w)$}}
\psfrag{QC}[bc][bc]{\small \ \ QCP}
\psfrag{GFL}[bc][bc]{\small GFL}
\psfrag{LM}[bc][bc]{\small LM}
\epsfig{file=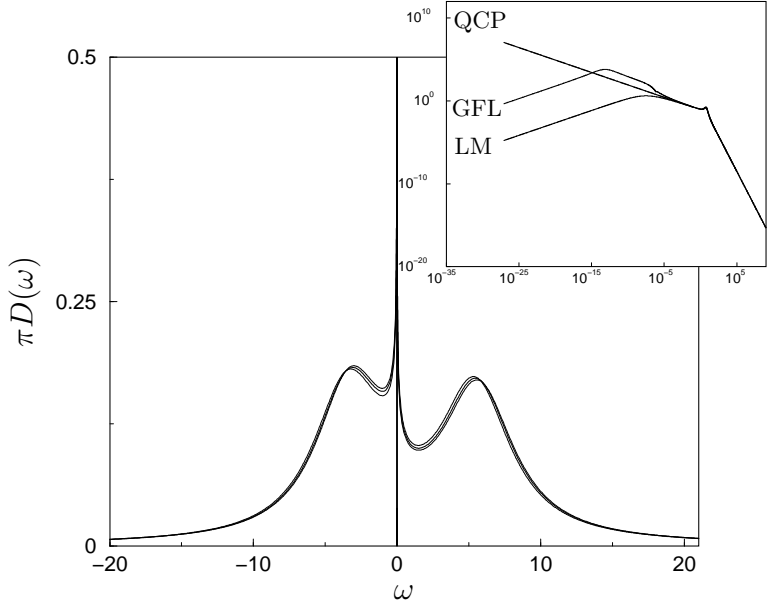,width=10cm}
\caption{Single-particle dynamics close to and at the QCP, for $r=0.3$
and $\eta =0.2$. Inset: $\pi D(\omega)$ \it vs \rm $|\omega|$  for $\omega>0$ on a
logarithmic scale, for $U=12.0$ (GFL), $12.4$ (LM) and $12.2$ (QCP). The main
figure shows the corresponding spectra \it vs \rm $\omega$ on a linear scale. Full
discussion in text.}
\label{}
\end{center}
\end{figure}

  As $U \rightarrow U_{\cc}(r,\eta)\pm$ and the low-energy scale $\omega_{*} \rightarrow
0$, this $|\omega|^{-r}$ power law persists to lower and lower frequencies, and
alone survives at the QCP where $\omega_{*}=0$ identically and the low-energy
physics is scale-free. The LMA enables a direct determination of the spectrum
precisely \it at \rm the QCP $U=U_{\cc}(r,\eta)$ (numerically in general,
and analytically for small $r$ as in \S 7). The resultant QCP spectrum for
$r=0.3, \eta =0.2$ is also shown in figure; from which the characteristic
behaviour $\pi D(U_{\cc};\omega) = C(r,\eta)|\omega|^{-r}$ as $|\omega| \rightarrow 0$
is clearly seen (and which in fact persists essentially up to non-universal
scales $\omega \sim {\cal{O}}(1)$).

 One obvious but important point should also be noted from the
above discussion: close to the transition, and in \it either \rm phase,
$\omega_{*}$ acts as the natural crossover frequency scale to the
emergence of \it common \rm QCP behaviour for $|\w|\gg \w_*$ in single-particle dynamics,
while for lower energies the spectral characteristics are particular to
the phase considered. This is of course the dynamical analogue, at $T=0$,
of the common schematic pertaining to static/thermodynamic properties at
finite-$T$ (see e.g.\ [21] or figure 1 of [19]), showing the crossover to QCP
behaviour upon increasing $T$ above $T \sim \omega_{*}$.

\subsection{Scaling: a closer look}
In the preceding sections we have focussed primarily on the asymmetric
PAIM, $\eta \neq 0$.
The purpose of this section is three-fold. First, to compare results
for $\eta \neq 0$ to those arising in the particle-hole symmetric limit
$\eta =0$ [15,16,20]. There are of course important differences between the two ---
notably that for $\eta =0$ the LM phase alone arises for all $r>\frac{1}{2}$ and
any $U\neq 0$ [10-20] --- but, that aside, these are not as radical as might perhaps
be inferred from the literature and the two may be readily understood on a common
footing.

\begin{figure}
\begin{center}
\psfrag{yaxis}[bc][bc]{{\large ${\cal F}(\w)$}}
\psfrag{xaxis}[bc][bc]{{\large  ${\tilde{\w}}$}}
\psfrag{a}[bc][bc]{\large (A)}
\epsfig{file=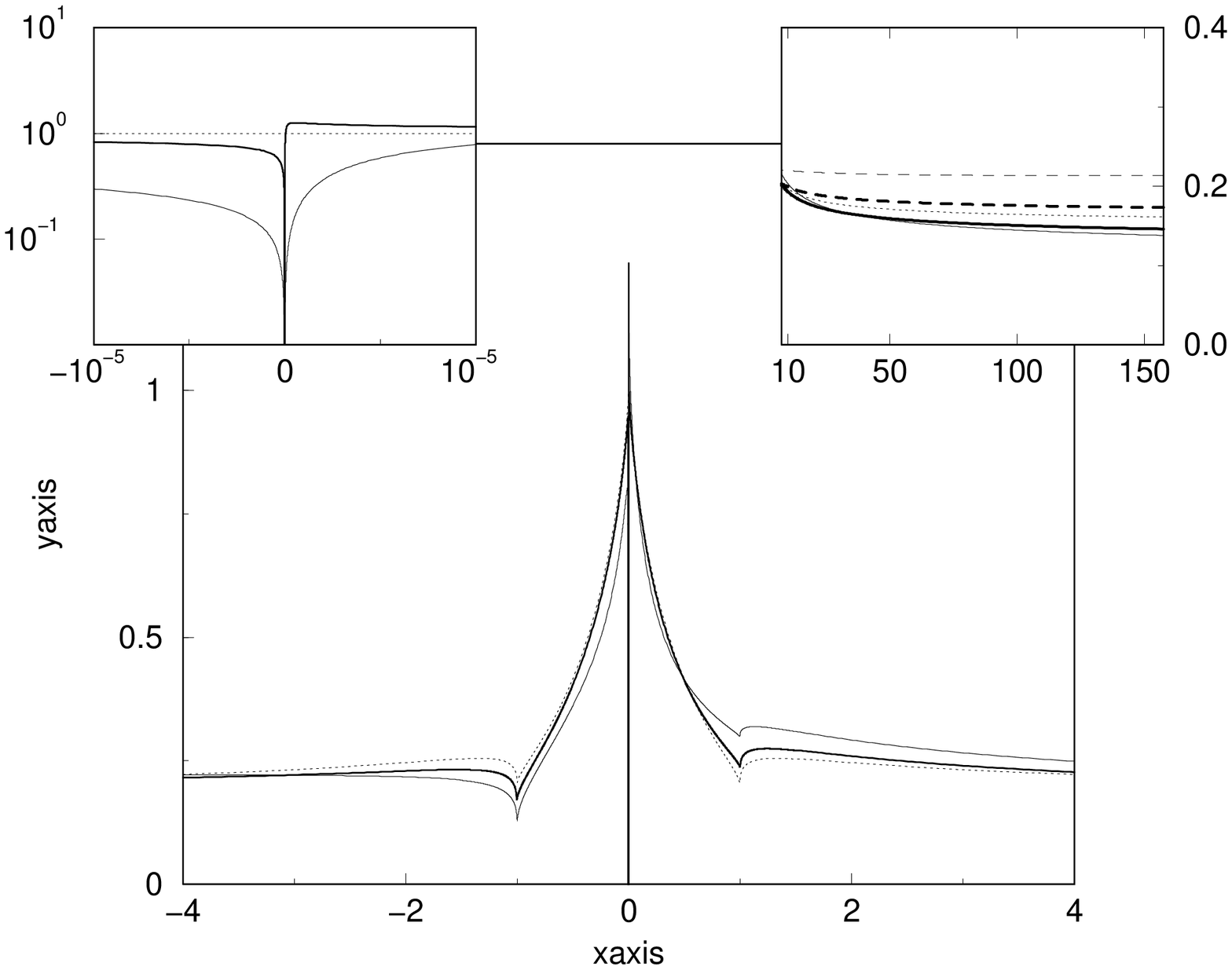,width=10cm}
\psfrag{yaxis}[bc][bc]{{\large $\pi\Delta_0\wl^r D(\w)$}}
\psfrag{xaxis}[bc][bc]{{\large  ${\tilde{\w}}$}}
\psfrag{b}[bc][bc]{\large (B)}
\epsfig{file=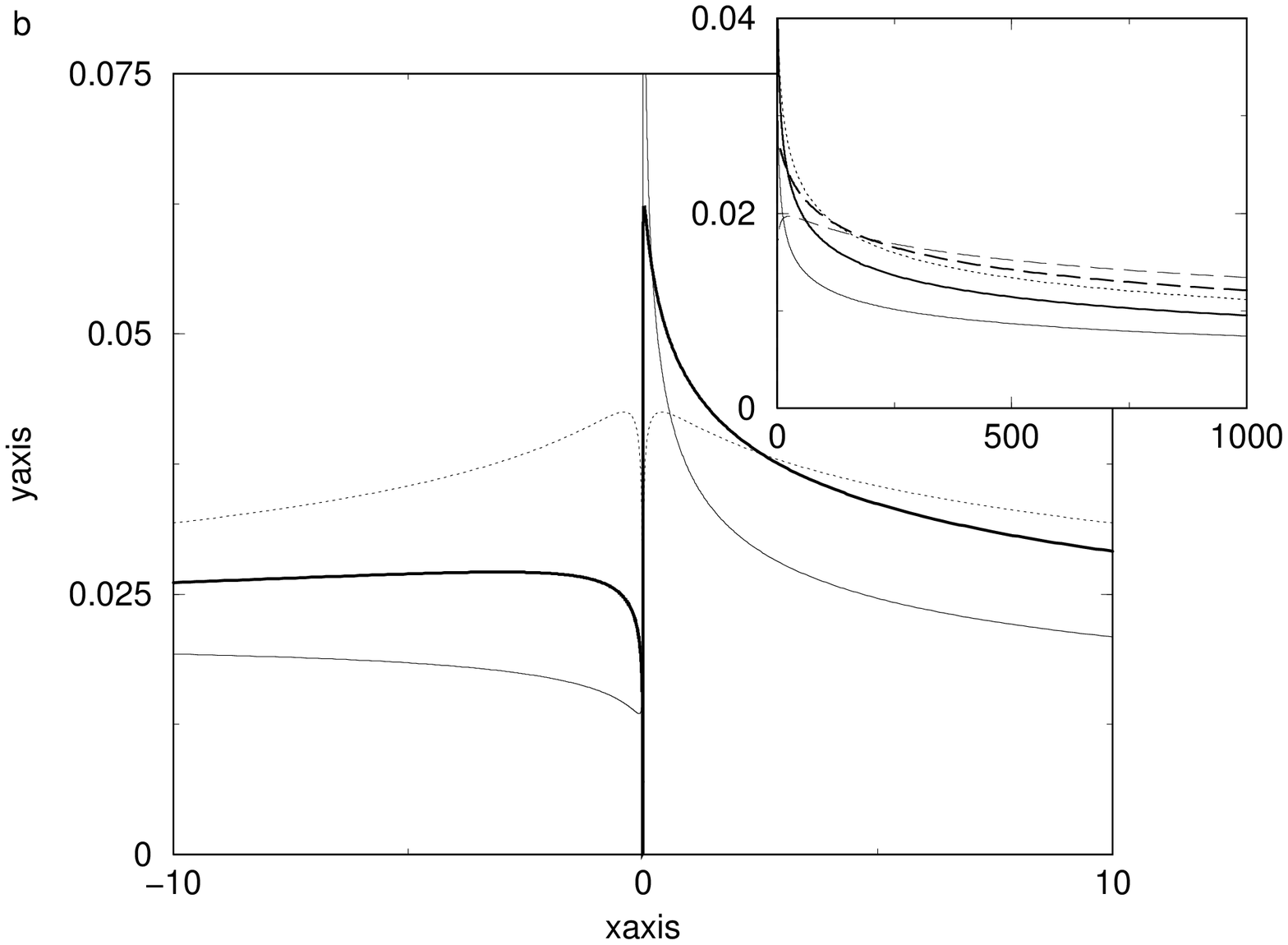,width=10cm}

\caption{
Scaling spectra in both phases for fixed $r$ ($=0.3$) and varying
asymmetry $\eta$. \bf A (upper): \rm GFL phase scaling spectrum ${\cal{F}}(\omega)
=$sec$^{2}(\frac{\pi}{2}r)|\tilde{\omega}|^{r} \times \pi\omega_{*}^{r}D(\omega)$
\it vs \rm $\tilde{\omega} =\omega/\omega_{*}$ ($\omega_{*} \equiv \wm$)
for asymmetries  $\eta =0$ (dotted line), 0.2 (bold solid) and $0.5$ (thin solid line).  The main figure shows the central `Kondo resonance'.  The right inset shows the  spectral `tails' out to $|\tilde{\omega}| = 150$, $\eta =0$ (dotted line), $\eta =0.2$ (bold solid line for $\omega>0$, bold
dashed line for $\omega <0$) and $\eta = 0.5$  (thin solid  [dashed] line for
$\omega >0$ [$\omega <0$]).  Left inset: scaling spectra
for $\eta =0, 0.2$ and $0.5$ on the lowest scales (\it cf \rm figure 11).
\bf B (lower): \rm LM phase scaling spectrum $\pi \omega_{*}^{r}D(\omega)$ \it vs \rm
$\tilde{\omega} =\omega/\omega_{*}$ ($\omega_{*} \equiv \wl$) for
$\eta =0$ (dotted line), $\eta =0.2$ (bold solid line) and $\eta =0.5$ (thin solid
line). Inset: spectra shown out to $|\tilde{\omega}| =10^{3}$ for both
$\tilde{\omega}>0$ and $<0$, same labelling as in figure 17A.}
\end{center}
\end{figure}

  In comparing dynamics for $\eta \neq 0$ and $\eta =0$, an obvious but important
point should first be noted: that we have considered the scaling spectra in
terms of $\tilde{\omega} =\omega/\omega_{*}$ and that the low-energy scale $\omega_{*}$
is non-vanishing throughout both phases, \it including the particle-hole
symmetric limit \rm $\eta =0$ [15,20]. In the GFL phase for example one might have chosen, as in \S 6.1, to express scaling in terms of $\omega/T^{*}$; with
$T^{*} \propto |\epsilon_{\ii}^{*}|^{\frac{1}{r}}$
given by equation (6.4) such that the low-energy maxima in the GFL phase spectra
lie at $\omega/T^{*} = \pm 1$. This is the procedure adopted in [18] and --- in
general --- is entirely equivalent to ($\omega/\omega_{*}$)-scaling because
$T^{*} \propto \omega_{*}$. But the proportionality between the two, given
(see equation (3.11)) from
$k(r,\eta) = \frac{|\epsilon_{\ii}^{*}|}{\omega_{*}^{r}} \sim
(\frac{T^{*}}{\omega_{*}})^{r}$ is odd in $\eta$, such that $\epsilon_{\ii}^{*}$
vanishes \it throughout \rm the GFL phase
in the symmetric limit, while $\omega_{*} \equiv \wm$ remains non-zero.
For this reason, scaling in terms of $\omega_{*}$ is naturally required to consider
the symmetric and asymmetric PAIMs on an equivalent footing.

In comparing the GFL phase scaling spectra for different $\eta$, it is also
convenient to consider the so-called modified spectral function [14-16,20]
${\cal{F}}(\omega) =$sec$^{2}(\mbox{$\frac{\pi}{2}$}r)\pi |\omega|^{r}D(\omega)$, i.e.
\begin{equation}
{\cal{F}}(\omega) = sec^{2}(\mbox{$\frac{\pi}{2}$})r) |\tilde{\omega}|^{r} \times
\pi \omega_{*}^{r}D(\omega)
\end{equation}
(likewise universally dependent on $\tilde{\omega}$).
For $\eta =0$ the resultant
spectrum is `pinned' at the Fermi level for all $U$ in the GFL phase [14-16],
${\cal{F}}(\omega=0) =1$ (as follows from equation (3.5)); while for any $\eta \neq 0$,
${\cal{F}}(\omega) \propto |\tilde{\omega}|^{2r}$ as $|\tilde{\omega}| \rightarrow 0$.
The scale on which this distinction between the $\eta =0$ and $\eta \neq 0$ scaling
spectra arises is $T^{*}$ ($\propto \omega_{*})$ as discussed in \S 6.1 (figure 11).
But in practice the ratio $T^{*}/\omega_{*}$ is found within the LMA to
be small ($T^{*}/\omega_{*}  \sim [k(r,\eta]^{\frac{1}{r}}$
with $k$ itself small compared to unity), so the qualitative differences between
$\eta =0$ and $\neq 0$ are apparent only on the lowest-$\tilde{\omega}$ scales.
This is seen clearly in the left inset to figure 17a which, for $r=0.3$ and asymmetries
$\eta =0, 0.2$ and $0.5$, shows ${\cal{F}}(\omega)$ \it vs \rm $\tilde{\omega}$
for $|\tilde{\omega}| \ll 1$.

  By contrast however, the strong similarities between particle-hole
symmetric and asymmetric spectra in the GFL phase are also evident in figure 17a.
The main figure gives ${\cal{F}}(\omega)$ on the $|\tilde{\omega}| \sim{\cal{O}}(1)$
scale, showing clearly the low-energy `Kondo resonance' in
${\cal{F}}(\omega)$ [14-16] (which becomes precisely the usual Kondo resonance of the
metallic AIM in the $r=0$ limit [14-16]). The effects of non-zero $\eta$,
while present, do not lead to qualitatively different behaviour either on the
$\tilde{\omega} \sim{\cal{O}}(1)$ scales or in the spectral `tails', shown
out to $\tilde{\omega} \sim {\cal{O}}(10^{2})$ in the right inset. These comments
hold also for the LM phase scaling spectra, $\pi \wl^{r}D(\omega)$
\it vs \rm $\tilde{\omega} = \omega/\wl$ being shown in figure 17b
(again for $r=0.3$ and $\eta =0, 0.2$ and $0.5$). In this case
$\pi \wl^{r}D(\omega) \propto |\tilde{\omega}|^{r}$ as
$\tilde{\omega} \rightarrow 0$ for both the symmetric and asymmetric cases.
The spectra for
different $\eta$ are thus qualitatively similar for \it all \rm $\tilde{\omega}$,
albeit that the extent of spectral asymmetry naturally increases with $\eta$.

 In previous sections we have emphasised in particular the crossover from the
$\propto |\tilde{\omega}|^{r}$ behaviour of $\pi \omega_{*}^{r}D(\omega)$ arising
at low-$\tilde{\omega}$ to the ultimate large-$\tilde{\omega}$ behaviour
$\propto |\tilde{\omega}|^{-r}$ that in turn controls the QCP spectrum. Our
second point is simply to note that for the GFL phase, the asymptotic
approach to the ultimate $|\tilde{\omega}|^{-r}$ form --- where the
corresponding ${\cal{F}}(\omega) \propto |\tilde{\omega}|^{r} \times
\omega_{*}^{r}D(\omega)$ thus plateaus to a constant --- is quite subtle. In
figure 17a for example, where the spectral tails of ${\cal{F}}(\omega)$ are
shown out to $|\tilde{\omega}| =150$, ${\cal{F}}(\omega)$ has not yet reached
its plateau value but is still decreasing, albeit slowly;
indeed from figure 12 it is clear the plateau does not occur
until $\tilde{\omega} \sim {\cal{O}}(10^{3})$ for the case of $r=0.3$ shown.
The approach to the ultimate large-$\tilde{\omega}$ behaviour of the scaling spectrum
is thus slow, and in fact becomes progessively more so with diminishing
$r \rightarrow 0$. The nature of this behaviour (which is relevant to the GFL phase
alone) is known from our recent work on
the symmetric PAIM [20] and is modified only quantitatively for $\eta \neq 0$:
as $r \rightarrow 0$, the ultimate large-$|\tilde{\omega}|$ behaviour arises only for
$|\tilde{\omega}| \gg e^{\frac{1}{r}}$. For
$1 \ll |\tilde{\omega}| \ll e^{\frac{1}{r}}$ by contrast the decay of
${\cal{F}}(\omega)$ towards its plateau value is logarithmically slow. This
`intermediate' large-$\tilde{\omega}$ regime progressively dominates the GFL
scaling spectrum out to ever larger-$\tilde{\omega}$ scales as $r \rightarrow 0$,
and for $r=0$ recovers precisely the
known [26] logarithmic large-$\tilde{\omega}$ tails characteristic of the
metallic AIM (for which $D(\omega) \propto 1/ln^{2}(|\tilde{\omega}|)$
for $|\tilde{\omega}| \gg 1$). While it is the small-$r$ behaviour to which
we turn in the final section, the reader is referred to [20] for further details
on this point.

 As noted above, the ultimate large-$\tilde{\omega}$ behaviour
$\omega_{*}^{r}D(\omega) \propto |\tilde{\omega}|^{-r}$ is not
in fact reached in the $\tilde{\omega}$-range shown in figure
17a for the GFL phase. In consequence \it none \rm of the behaviour
illustrated in the figure survives at the QCP itself. Our final remark
here is simply to emphasise this point, evident also in figure 16 (inset):
the Kondo resonance in ${\cal{F}}(\omega)$, illustrated clearly figure 17a
and characteristic of the GFL phase, `collapses' at the QCP
where its weight vanishes; as indeed remarked in [19], and which the above
figures substantiate.

\seceq

\section{Small $r$ asymptotics}

Our aim now is to obtain analytical results for the critical behaviour of the
low-energy scales $\omega_{*}$, the phase boundaries and single-particle scaling
spectra, in the strong coupling regime where $U_{\cc}(r,\eta) \gg 1$. This means
in effect that we consider here small $r \ll 1$, since $U_{\cc}^{-1} \propto r$ as
$r \rightarrow 0$ (see e.g.\ figure 8).

   In strong coupling, where the impurity charge $n \rightarrow 1$, the
low-energy behaviour of the pseudogap AIM maps under a Schrieffer-Wolff
transformation onto the corresponding Kondo model [2], with host density of
states $\rho_{\mbox{\ssz{host}}}(\omega) = \rho_{0}|\omega|^{r}\theta (D-|\omega|)$.
The  exchange coupling matrix element
$J =2V^{2}[1/|\epsilon_{\ii}|+1/(U-|\epsilon_{\ii}|)]$, i.e.
$J = 8V^{2}/[U(1-\eta^{2})]$ expressed in terms of $U$ and $\eta \equiv
1-2|\epsilon_{\ii}|/U$; and the corresponding potential scattering strength is $K=\eta J$. Since the PAIM hybridization $\Delta_{\subi}(\omega) =
\pi V^{2}\rho_{\mbox{\ssz{host}}}(\omega)$ (with $\Delta_{\subi}(\omega)$ given by equation (2.4)),
it follows that
\begin{equation}
\rho_{0}J = \frac{8}{\pi U} \frac{1}{1-\eta^{2}}
\end{equation}
from which strong coupling PAIM results can be transcribed directly to those
applicable to the Kondo model.

 As a straightforward generalisation of [15,20] we first focus on
$\{\Sigma_{\sigma}(\omega) \}$ in the GFL phase (equation (4.14)),
where in strong coupling $\Sigma_{\sigma}(\omega)$ takes the form
\begin{equation}
\Sigma_{\sigma}(\omega) \sim U^{2}{\cal{G}}_{-\sigma}^{-\sigma}(\omega
+\sigma\wm).
\end{equation}
This follows from the fact that the spectral weight of Im$\Pi^{+-}(\omega)$ in
strong coupling is confined asymptotically to $\omega >0$ with
$\int^{\infty}_{0}(\rmd\omega/\pi)$Im$\Pi^{+-}(\omega) =1$ (which physically
reflects saturation of the local moment, $|\mu| \rightarrow 1$); and that its
resonance is centred on the low-energy spin-flip scale $\omega_{m}$, on which scales
the $\{{\cal{G}}^{\sigma}_{\sigma}(\omega) \}$ are slowly varying. From equation (4.14),
equation (7.2) thus results. Equation (7.2) with $\wm=0$ also holds for the
LM phase, with $\wm =0$ reflecting as ever the local spin-degeneracy
inherent to the LM state; as follows from equation (4.18) noting that
[15,20] the LM poleweight $Q \rightarrow 1$ in strong coupling (and that the
$^{\s}\Sigma_{\sigma}(\omega)$ contribution is both negligible in intensity and
subdominant in frequency [15]).

  Re${\cal{G}}^{\sigma}_{\sigma}(\omega)$ is given by the one-sided Hilbert
transform equation (4.15), and its leading low-$\omega$ behaviour in strong coupling (and
hence for $r \rightarrow 0$) is readily obtained using the methods of [15,20,27].
The result is
\begin{equation}
U^{2}\mbox{Re}{\cal{G}}_{-\sigma}^{-\sigma}(\omega) \sim
\sigma\frac{4\lambda^{r}}{\pi r} g_{\sigma}(e_{\ii}^{\prime})
\left[1-\frac{|\omega|^{r}}{\lambda^{r}}\right] +{\cal{O}}(|\omega|)
\end{equation}
where $\lambda =$min$(D,U)$ arises as a high-energy cutoff [15,20,27] and
$g_{\sigma} \equiv g_{\sigma}(e_{\ii}^{\prime}) = [1+\sigma e_{\ii}^{\prime}]^{-2}$,
with $e_{\ii}^{\prime} =2e_{\ii}/U$ and $e_{\ii}$ determined as specified in \S s 4.1.1
and 4.3. Equations (7.2) and (7.3) for $\Sigma^{\subr}_{\sigma}(\omega =0)$
will now be employed to determine the critical behaviour of $\omega_{*} \equiv
\wm$, and consequent phase boundaries.
We also note that the introduction of $\lambda =$min$(D,U)$ means the following
analysis encompasses not only $D \gg U$ (as exemplified by the wide band limit
$D=\infty$ employed explicitly in the full calculations of
previous sections), but in addition the case $U \gg D$.
As will be seen, the essential differences between the two naturally reside only in the
precise dependence of the low-energy scales and phase boundaries [15,16] on bare model
parameters; while for obvious physical reasons the scaling spectra for
both phases are entirely independent of whether $D \gg U$ or \it vice versa\rm.

\subsection{Statics: scales and phase boundaries}
We consider first the approach to the transition from the GFL phase,
determining thereby the critical $U_{\cc}(r,\eta)$ for small-$r$ and the
critical behaviour of the Kondo spin-flip scale $\wm \equiv \omega_{*}$
as $U \rightarrow U_{\cc}-$.
A brief discussion is then given of resultant phase boundaries, and the corresponding
low-energy scale $\omega_{*} \equiv \wl$ characteristic of the LM phase.

The symmetry restoration condition equation (4.5) is a central element of our approach
to the GFL phase. Using equation (4.9) it is given in strong coupling (where $|\bar{\mu}|
\rightarrow 1$) by $\Sigma^{\subr}_{\uparrow}(0)-\Sigma^{\subr}_{\downarrow}(0) = U$; and
hence from equations (7.2,3) by
\begin{equation}
\frac{4\lambda^{r}}{\pi r} \left[1 - \left(\frac{\wm}{\lambda}\right)^{r}\right]
{\sum_{\sigma}}g_{\sigma} = U.
\end{equation}
From this the critical $U_{\cc}(r,\eta)$, where $\wm =0$, follows directly
from $(\lambda^{r}/U_{\cc}(r,\eta)) {\stackrel{r \rightarrow 0}{\sim}}
(\pi r/4)({\sum_{\sigma}}g_{\sigma})^{-1}$, whence $U_{\cc}^{-1}$ vanishes linearly
in $r$ as $r \rightarrow 0$ as noted in the discussion of figure 8 (\S 5.1).
In the particle-hole symmetric ($\eta =0$) limit, $e_{\ii}^{\prime} =0$ (for all $r$ and
$U$, see \S 4) and hence $g_{+} =1 = g_{-}$; the result obtained previously in
[15] is then recovered, $U_{\cc}^{-1}(r,\eta =0) = \pi r/8$ as $r \rightarrow 0$,
which is exact [15,16]. More
generally a knowledge of the $g_{\sigma}$ as $r \rightarrow 0$ is required. But since
$U_{\cc}^{-1}(r,\eta) \propto r$ as above, the $e_{\ii}^{\prime}$ and hence $g_{\sigma}$
may be replaced
asymptotically by their corresponding $r=0$ values [27]; these are $U$-independent
and functions solely of the asymmetry $\eta$, given explicitly within the present
LMA by [27]
\begin{equation}
g_{\pm} = \frac{f(\eta^{2})}{1 \pm \eta} \ \ \ \ : \ \ f(\eta^{2})
=\mbox{$\frac{1}{2}$} [1 + (1-\eta^{2})^{\frac{1}{2}}].
\end{equation}
(This result may also be obtained by detailed consideration of the limit
$r \rightarrow 0$ itself, analysis of the renormalised level $\epsilon_{\ii}^{*}
\equiv \epsilon_{\ii} + \tilde{\Sigma}^{\subr}_{\sigma}(0)$ showing both that
the $g_{\sigma}$ given above hold as $U \rightarrow U_{\cc}(r,\eta)$
and that the exponent for the vanishing of $|\epsilon_{\ii}^{*}| \sim u^{a_{1}}$
(equation (3.10)) is indeed $a_{1} =1$ as in figure 4.) From equations (7.4,5) the
critical $U_{\cc}(r,\eta)$ is thus given from
\begin{equation}
\frac{\lambda^{r}}{U_{\cc}(r,\eta)} \stackrel{r \rightarrow 0}{\sim}
\frac{\pi r}{8} \frac{(1-\eta^{2})}{f(\eta^{2})}
\end{equation}
or equivalently
\begin{equation}
f(\eta^{2}) \rho_{0}J_{\cc} \stackrel{r \rightarrow 0}{\sim}
\frac{r}{\lambda^{r}}
\end{equation}
for the critical exchange coupling of the corresponding Kondo model.

 We shall return to this below, but note first that the
$U \rightarrow U_{\cc}(r,\eta)-$  critical behaviour
of the Kondo scale $\wm$ follows in turn from equation (7.4) as
\begin{equation}
\frac{\wm}{\lambda} \sim
\left(1-{\frac{U}{U_{\cc}(r,\eta)}}\right)^{\frac{1}{r}}.
\end{equation}
The exponent for the vanishing of the
low-energy GFL scale $\wm \equiv \omega_{*}$ is thus $1/r$; as found
in figure 3 where, with $\lambda =$min$(D,U) \equiv U$ as appropriate to the
wide band limit $D = \infty$ there employed, $(\omega_{*}/U)^{r}$ \it vs \rm
$U$ was shown to vanish linearly in $u = (1-U/U_{\cc})$ as $u \rightarrow 0+$.
We add moreover that using equation (7.6) in equation (7.8) and taking the limit
$r \rightarrow 0$ recovers precisely the Kondo scale hitherto found in [27]
for the $r=0$ metallic AIM,
\begin{equation}
\frac{\wm(r=0)}{\lambda} = exp\left(-\frac{\pi U}{8} \frac{(1-\eta^{2})}
{f(\eta^{2})} \right ) \equiv exp\left( -\frac{1}{\rho_{0}J}\frac{1}{f(\eta^{2})}
\right )
\end{equation}
(with $J \equiv J(U,\eta^{2})$ given by equation (7.1)). While recovery of an exponentially
small Kondo scale from an approximate theory is non-trivial, we note that the exponent
therein differs in general from the exact result [2] for the $r=0$ Kondo model by the
factor of $f(\eta^{2}) \in [1,\frac{1}{2}]$. It is thus as such exact only for the
symmetric model where $f(\eta =0) =1$; albeit that the correction is modest, $f$
being slowly varying in $\eta$ and lying e.g. within $10\%$ of unity for
$\eta <0.6$. Since equation (7.6) (or equivalently equation (7.7)) likewise reflects the
$r=0$ exponent, the $\eta$-dependence of the corresponding exact result as
$r \rightarrow 0$ is obtained by replacing $f(\eta^{2})$ by $1$ therein.

 The $r \rightarrow 0$ phase boundary arising from equation (7.6)
is shown in figure  6 of \S 5.2 (with $\lambda \equiv U$ as appropriate to
the wide-band limit), where it is compared to full numerical results for
$r=0.2$ and displayed as $\eta$ \it vs \rm
$(8/\pi r)U^{r-1}_{\cc}(r,\eta)$  --- which we note is equivalently
lim$_{r \rightarrow 0} [U_{\cc}(r,\eta=0)/U_{\cc}(r,\eta)]$.
Here we add in passing that the resultant $r \rightarrow 0$
phase boundary between non-magnetic (GFL) and magnetic (LM) states, is redolent
of the phase diagram obtained in Anderson's original paper on the $r=0$ metallic
AIM (figure 4 of [1]).
There is of course no transition for $r=0$: the `phase
diagram' of [1] was simply the boundary to local moment formation at MF level,
the qualitative deficiencies of which are long since known; in particular
that a spurious transition arises at a critical $U^{\mbox{\scriptsize{MF}}}_{\cc}(r=0,\eta)$
(with $U^{\mbox{\ssz{MF}}}_{\cc}(0,0) =\pi$). But the similarities between
the two are nonetheless striking and the $r=0$ MF results for
$\eta$ \it vs \rm $[U^{\mbox{\ssz{MF}}}_{\cc}(0,\eta =0)/U^{\mbox{\ssz{MF}}}_{\cc}(0,\eta)]$ ([1], figure 4),
while wrong for $r=0$ itself, appear ironically to be qualitatively correct
as $r \rightarrow 0$.

\begin{figure}
\begin{center}
\psfrag{xaxis}[bc][bc]{\large $\ei$}
\psfrag{yaxis}[bc][bc]{\large $U$}
\psfrag{lm}[bc][bc]{\Large LM}
\psfrag{gfl}[bc][bc]{\Large GFL}
\epsfig{file=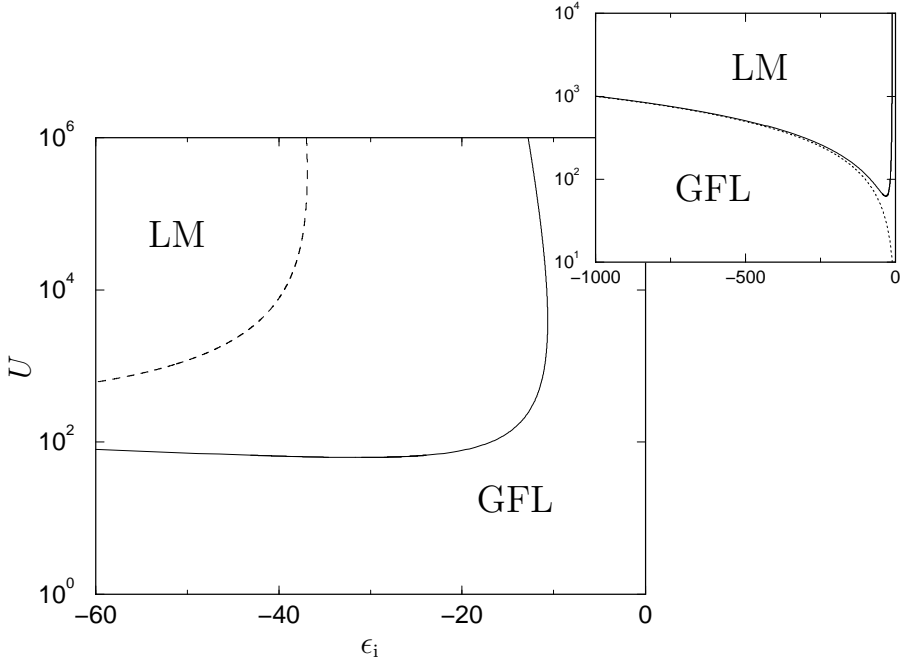,width=12cm}
\caption{Phase boundary in the $(U,\epsilon_{\ii})$-plane for fixed
$r=0.05$ (solid line), obtained from equation (7.6) (with $\lambda =U$ as appropriate to the wide band case). The corresponding result for $r=0.01$ is shown for comparison (dashed line). Inset: $r=0.05$ phase boundary for large negative $\epsilon_{\ii}$, which is seen to approach asymptotically $U_{\cc}=|\epsilon_{\ii}|+$.}
\label{pb4}
\end{center}
\end{figure}

  Figure 18 by contrast shows the phase diagram in the ($U,\epsilon_{\ii}$)-plane
arising from equation (7.6) (with $\lambda =U$), for both $r=0.05$ (solid line) and
$r=0.01$; the former showing again the re-entrant behaviour discussed
in \S 5.2 in connection with figure 5. The increasing stability of the GFL
phase with diminishing $r$ is clearly evident, and expected physically since
the Fermi liquid phase alone persists throughout the entire $(U,\epsilon_{\ii})$-plane
in the $r=0$ limit of the metallic AIM. The inset to the figure also shows
the phase boundary (for $r=0.05$) out to large negative $\epsilon_{\ii}$,
from which it is seen that the critical $U_{\cc}$ asymptotically approaches
$|\epsilon_{\ii}|+$ (shown as a dotted line) --- again physically natural, this
being the condition for local moment formation in the atomic limit.

 In figure 19 the $r$-dependence of full LMA phase boundaries is shown for a
range of asymmetries $\eta$, and the wide-band limit employed in previous sections.
Specifically, $F(r,\eta) =
(8/\pi)[f(\eta^{2})/(1-\eta^{2})]U_{\cc}^{r-1}(r,\eta)$ \it vs \rm $r$ is shown,
which from equation (7.6) should behave as $F(r,\eta) \sim r$ as $r \rightarrow 0$
such that phase boundaries for different $r$ then collapse to a common curve.
This behaviour is indeed seen in figure 19 and we note that, in line with results
from a previous NRG study [13], the resultant phase boundaries have little
asymmetry dependence up to $r \simeq 0.25$. The asymptotic behaviour
$F(r,\eta) \sim r$ is also shown on the figure and seen to be reached in
practice below $r \sim 0.1$ or so (which range we add is known from a previous
LMA and NRG study [16] of the symmetric limit to be enhanced
further if one works instead with a finite bandwith $D$, with $U/D \gg 1$).

\begin{figure}
\begin{center}
\psfrag{xaxis}[bc][bc]{{\large  $r$}}
\psfrag{yaxis}[bc][bc]{{\large  $F(r,\eta)$}}

\epsfig{file=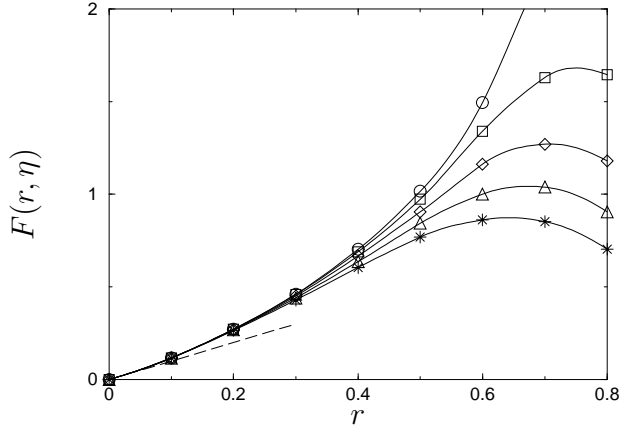,width=8cm}
\caption{Full LMA phase boundaries, shown as $F(r,\eta) \equiv
(8/\pi)[f(\eta^{2})/(1-\eta^{2})]U^{r-1}$ \it vs \rm $r$, for different asymmetries $\eta =0.1$, 0.2, 0.3, 0.4 and $0.5$ (from top to bottom in sequence). As
$r \rightarrow 0$, all collapse to the common form $F(r,\eta) \sim r$ (itself
shown, dashed line).}
\label{pb6}
\end{center}
\end{figure}

  We now consider briefly the corresponding $U \rightarrow U_{\cc}+$ critical
behaviour of the low-energy scale $\omega_{*} \equiv \wl$ characteristic
of the LM phase, given by equation (6.6) (the resultant
phase boundary $U_{\cc}(r,\eta)$ is of course correctly independent of the phase
from which it is approached). This is straightforward, since in strong coupling
(where $|\bar{\mu}| \rightarrow 1$), $\tilde{\Sigma}^{\subr}_{\uparrow}(0) -
\tilde{\Sigma}^{\subr}_{\downarrow}(0) = \Sigma_{\uparrow}^{\subr}(0) -
\Sigma_{\downarrow}^{\subr}(0) -U$ from equation (4.9); and $\Sigma_{\sigma}^{\subr}(0)$
is given from equations (7.2,3) with $\wm =0$ throughout the LM phase,
whence $\tilde{\Sigma}^{\subr}_{\uparrow}(0) -\tilde{\Sigma}^{\subr}_{\downarrow}(0)
= (4\lambda^{r}/\pi r)\sum_{\sigma}g_{\sigma} -U$. But from equation (7.4)
$(4\lambda^{r}/\pi r)\sum_{\sigma}g_{\sigma} = U_{\cc}$, so
$\tilde{\Sigma}^{\subr}_{\uparrow}(0) -\tilde{\Sigma}^{\subr}_{\downarrow}(0)
\sim -U_{\cc}(U/U_{\cc} -1)$ as $U \rightarrow U_{\cc}+$, with
$U_{\cc} \propto 1/r$ as $r \rightarrow 0$ from equation (7.6). Hence
$\wl \propto (r|\tilde{\Sigma}^{\subr}_{\uparrow}(0) -
\tilde{\Sigma}^{\subr}_{\downarrow}(0)|)^{1/r}$ given by equation (6.6) indeed has the
critical behaviour
\begin{equation}
\wl \propto \left( \frac{U}{U_{\cc}(r,\eta)} -1\right)^{\frac{1}{r}}
\end{equation}
with exponent $1/r$, as shown numerically in figure 3 (\S 5.1).

\subsection{Scaling spectra and the QCP}
 We turn finally to the scaling spectra characteristic of the GFL and LM phases
(figures 12, 15), in particular the leading large $\tilde{\omega} = \omega/\omega_{*}$
behaviour thereof that determines the spectrum at the quantum critical point
itself (\S 6.3).

The impurity Green function is given by equations (2.9,10), which may be expressed
quite generally as $G(\omega) = \mbox{$\frac{1}{2}$}\sum_{\sigma} [\omega^{+} - \Delta(\omega)
- \epsilon_{\ii\sigma}^{*} -(\Sigma_{\sigma}(\omega) - \Sigma_{\sigma}(0))]^{-1}$,
where we use $\tilde{\Sigma}_{\sigma}(\omega) -
\tilde{\Sigma}_{\sigma}(0) = \Sigma_{\sigma}(\omega) -\Sigma_{\sigma}(0)$
(trivially, from equation (4.9)). Here as usual,
$\epsilon_{\ii\sigma}^{*} = \epsilon_{\ii} + \tilde{\Sigma}_{\sigma}(0)$ is the renormalised level (equation (5.20)); with the $\epsilon_{\ii\sigma}^{*} = \epsilon_{\ii}^{*}$
independent of spin $\sigma$ for the GFL phase where symmetry is restored (equation (4.5)),
but $\epsilon_{\uparrow}^{*} \neq \epsilon_{\downarrow}^{*}$ for the LM phase, as
shown in \S 5.1. As discussed in the preceding sections, (i) the scaling spectra for
the GFL or LM phases are obtained by considering finite
$\tilde{\omega} =\omega/\omega_{*}$ (with $\omega_{*} = \wm$ or $\wl$
as appropriate), in the formal limit $\omega_{*} \rightarrow 0$; and (ii) it is
$\omega_{*}^{r}G(\omega)$ (as opposed to $G(\omega)$ itself) which exhibits universal
scaling in terms of $\tilde{\omega}$. This follows directly,
$\omega_{*}^{r}G(\omega) =
\mbox{$\frac{1}{2}$}\sum_{\sigma}[\omega_{*}^{1-r}\tilde{\omega}^{+} -
\omega_{*}^{-r}\Delta(\omega) - \omega_{*}^{-r}(\epsilon_{\ii\sigma}^{*} +
(\Sigma_{\sigma}(\omega) - \Sigma_{\sigma}(0))]^{-1}$, in which
$\omega_{*}^{1-r}|\tilde{\omega}|$ may be neglected entirely for $0 \leq r <1$
as considered.
By contrast, from equations (2.4,5) for the hybridization, $\omega_{*}^{-r}\Delta(\omega)$
is non-vanishing in the scaling regime and given by $\omega_{*}^{-r}\Delta(\omega)
= -$sgn$(\omega)[\beta(r)+\im]|\tilde{\omega}|^{r}$ (independently of the bandwith
$D$). The scaling spectrum $\pi \omega_{*}^{r}D(\omega) =
-$sgn$(\omega)\omega_{*}^{r}$Im$G(\omega)$ is thus obtained from
\begin{equation}
\fl\omega_{*}^{r}G(\omega) = {\mbox{$\frac{1}{2}$}}{\sum_{\sigma}}\left[\mbox{sgn}(\omega)[\beta (r)+\im]|\tilde{\omega}|^{r} - \frac{\epsilon_{\ii\sigma}^{*}}
{\omega_{*}^{r}} - \omega_{*}^{-r}(\Sigma_{\sigma}(\omega) -\Sigma_{\sigma}(0))
\right]^{-1}
\end{equation}
which we add holds generally for all $0 \leq r < 1$ considered; and where
$\epsilon_{\ii\sigma}^{*}/\omega_{*}^{r}$ as $\omega_{*} \rightarrow 0$ has been
shown in \S 5.1 to tend to a non-zero constant for both the GFL and LM phases
(denoted therein as $k(r,\eta)$ for the GFL phase).

Since $\omega_{*}^{r}G(\omega)$ scales, it follows from equation (7.11) that
$\omega_{*}^{-r}[\Sigma_{\sigma}(\omega) - \Sigma_{\sigma}(0)]$ is likewise
universally dependent solely on $\tilde{\omega} = \omega/\omega_{*}$. In
the strong coupling, $r \rightarrow 0$ regime of interest here, this may be
shown directly using equations (7.2,3) (together for $\Sigma^{\subi}_{\sigma}(\omega)$
with Im${\cal{G}}_{\sigma}^{\pm}(\omega) = -$sgn$(\omega)\pi
D_{\sigma}^{0}(\omega)\theta (\pm\omega)$ with $D_{\sigma}^{0}(\omega)$ the
MF spectral density). For the GFL phase one thereby finds
\alpheqn
\begin{equation}
\omega_{*}^{-r}[\Sigma^{\subr}_{\sigma}(\omega) - \Sigma^{\subr}_{\sigma}(0)] =
-\sigma {\frac{4}{\pi r}g_{\sigma}} \ \left[ |\tilde{\omega}+\sigma|^{r} -1 \right]
\end{equation}
\begin{equation}
\omega_{*}^{-r}\Sigma^{\subi}_{\sigma}(\omega) =
4g_{\sigma} \ \theta(-[1+\sigma\tilde{\omega}]) |\tilde{\omega} + \sigma|^{r}
\end{equation}
\reseteqn
(noting that $\Sigma^{\subi}_{\sigma}(0) =0$);
where the $g_{\sigma} \equiv g_{\sigma}(\eta)$ are given explicitly by equation (7.5)
and $\omega_{*} \equiv \wm$. For the LM phase by contrast,
\alpheqn
\begin{equation}
\omega_{*}^{-r}[\Sigma^{\subr}_{\sigma}(\omega) - \Sigma^{\subr}_{\sigma}(0)] =
-\sigma{\frac{4}{\pi r}}g_{\sigma}\ |\tilde{\omega}|^{r}
\end{equation}
\begin{equation}
\omega_{*}^{-r}\Sigma^{\subi}_{\sigma}(\omega) = 4g_{\sigma} \
\theta (-\sigma\tilde{\omega})|\tilde{\omega}|^{r}
\end{equation}
\reseteqn
with $\omega_{*} \equiv \wl$.
Equations (7.12,13), which we emphasise hold for all $\tilde{\omega}$
(and $r \rightarrow 0$), are indeed of the required universal form; and
for $\eta =0$, where the $g_{\sigma} =1$, reduce to the results of [20] for the
symmetric PAIM. Together with equation (7.11) they determine the full
$\tilde{\omega}$-dependence of the scaling spectra for the respective phases.
The resultant range of behaviour is rich, encompassing for example the
intermediate logarithmic regime in the GFL phase mentioned in \S 6.4, and
may again be obtained following the arguments of [20].

Here we consider solely
the ultimate large-$\tilde{\omega}$ behaviour of the scaling spectra for either
phase, relevant to the QCP itself. This arises for $|\tilde{\omega}|^{r} \gg 1$,
and for $r \rightarrow 0$ (where $\beta (r) =$tan$(\frac{\pi}{2}r) \sim 0$)
equations (7.11--13) yield the requisite behaviour
\begin{equation}
\fl \pi\omega_{*}^{r}D(\omega) \ \stackrel{\tilde{\omega}^{r} \gg 1}{\sim} \
\frac{\pi^{2}r^{2}}{32} \left (\frac{1}{g_{+}^{2}} + \frac{1+4g_{-}}{g_{-}^{2}} \right)
\ |\tilde{\omega}|^{-r} \ \equiv \ C(r,\eta)|\tilde{\omega}|^{-r}
\end{equation}
given explicitly for positive $\tilde{\omega}$ (for $\tilde{\omega} <0$, $g_{+}$ and
$g_{-}$ are simply interchanged). The result equation (7.14) applies to both the GFL
and LM phases; and $\omega_{*}^{r}$ cancels out of each side of the equation
to give the behaviour $\pi D(\omega) =C(r,\eta) |\omega|^{-r}$ precisely at the QCP
(which in that case extends down to $\omega =0$ (\S 6.3)). As found numerically
the QCP coefficient $C(r,\eta)$ is indeed both $r$- and $\eta$-dependent; with
$C(r,\eta) \sim r^{2}$ as $r \rightarrow 0$ and as such from equation (7.6) proportional
to $1/[U_{\cc}(r,\eta)]^{2}$ (or equivalently $[\rho_{0}J_{\cc}]^{2}$), in which the
interacting nature of the QCP is directly manifest.

\section{Summary}
 We have developed in this paper a non-perturbative local moment
approach to the $T=0$
pseudogap Anderson impurity model, with a conduction electron density
of states proportional to $|\omega|^{r}$ and itself a longstanding
paradigm [6-20] for understanding the topical and important issue of local
quantum
phase transitions.
The transition here is between a generalised fermi liquid phase,
perturbatively connected to the non-interacting limit, in which the impurity
spectrum exhibits a Kondo-like resonance
indicative of underlying local singlet behaviour; and a doubly degenerate
local moment phase in which the impurity spin remains unquenched. In addition
to
phase boundaries, and associated critical behaviour of relevant low-energy
scales,
local single-particle dynamics in the immediate vicinity of the transition
have been considered: in the GFL and LM phases separately,
and at the quantum critical point itself where the Kondo resonance has just
collapsed and the problem is scale-free.
Notwithstanding the range of vanishing low-energy scales as the transition
is approached, and their importance in understanding local spectra, dynamics
close to the QCP in either phase are nonetheless controlled by a single
low-energy scale $\omega_{*}$ and as such scale universally in terms of
$\omega/\omega_{*}$. A determination and understanding of these scaling
dynamics on \it all \rm $\omega/\omega_{*}$ scales, including their continuous
evolution `from the tails' to pure QCP behaviour, has been a primary focus of
the work. Approximate though it is the LMA is to our knowledge one of the
few theoretical approaches currently capable of handling these issues; and
the description arising from it is both rich and arguably
a good deal more subtle than appears to have been appreciated hitherto. Much
of course remains to be understood, not least the question of dynamics
at finite-temperature, aspects of which will be considered in a forthcoming
paper.

\ack
 We are grateful to the many people with whom we have had fruitful discussions
regarding the present work. Particular thanks are due to R Bulla, K Ingersent,
T Pruschke and M Vojta. We also thank the Leverhulme Trust, EPSRC and Balliol
College, Oxford, for support.

\end{document}